\documentstyle[12pt]{article}

\begin{document}

\begin{flushright}
IMSc/2015/06/04 
\end{flushright} 


\begin{center}

{\large \bf On the Stability of (M theory) Stars }

\vspace{4ex}

{\large \bf against Collapse : Role of Anisotropic Pressures}

\vspace{8ex}

{\large  S. Kalyana Rama}

\vspace{3ex}

Institute of Mathematical Sciences, C. I. T. Campus, 

Tharamani, CHENNAI 600 113, India. 

\vspace{1ex}

email: krama@imsc.res.in \\ 

\end{center}

\vspace{6ex}

\centerline{ABSTRACT}

\begin{quote} 

Unitarity of evolution in gravitational collapses implies
existence of macroscopic stable horizonless objects. With such
objects in mind, we study the effects of anisotropy of pressures
on the stability of stars. We consider stars in four or higher
dimensions and also stars in M theory made up of (intersecting)
branes. Taking the stars to be static, spherically symmetric and
the equations of state to be linear, we study `singular
solutions' and the asymptotic perturbations around them.
Oscillatory perturbations are likely to imply instability. We
find that non oscillatory perturbations, which may imply
stability, are possible if an appropriate amount of anisotropy
is present. This result suggests that it may be possible to have
stable horizonless objects in four or any higher dimensions, and
that anisotropic pressures may play a crucial role in ensuring
their stability.

\vspace{2ex}


\vspace{2ex}


\end{quote}

%
%

%
%
%
%

%
%
%

\newpage

\vspace{4ex}

\centerline{\bf 1. Introduction}  

\vspace{2ex}

A sufficiently massive star in four dimensional spacetime is
believed to be unstable against gravitational collapse.
Depending on its mass and its evolutionary stage, the star may
collapse to form a white dwarf or a neutron star or, if
supermassive, form a black hole. White dwarfs or neutron stars
may collapse further if they gain sufficient mass, for example,
by accretion. Thus, all sufficiently massive objects are
expected to collapse to ultimately form black holes \cite
{shapiro}. The same is expected in higher dimensional spacetime
also.

Assume that a black hole is formed in a collapse. It has a
horizon, emits Hawking radiation, and is believed to evolve
unitarily.  Presence of horizon means that no information from
inside the horizon is accessible to an outside observer. But
unitary evolution means that inside information must become
accessible to an outside observer atleast after some time.
Horizon must then cease to exist from this moment of
accessibility. The black hole at this time should still be of
macroscopic size so that information density within is atmost of
order Planckian scale, and not parametrically larger \cite {sbg,
hs}. And, the time when the horizon ceases to exist is expected
to be the Page time which is of the order of evaporation time
when about half the black hole has evaporated \cite {page93, 
page2, page3, sen}.

Another scenario is also possible where black holes do not form
at all. The gravitational collapse, that would have led to a
black hole, will instead lead to an object which has no horizon
as, for example, in Mathur's fuzz ball proposal \cite {fuzz,
fuzz2, fuzz3}. Unitary evolution in this scenario will proceed
as for any system with large number of degrees of freedom.

Thus, whichever scenario proves to be correct, the unitarity of
evolution implies that there must exist horizonless objects
which form about half way through a black hole evaporation if a
black hole has initially formed in a collapse, or right after a
gravitational collapse in which a black hole might have been
expected to form.

It must then be possible to construct such horizonless objects
using appropriate sources just as, for example, one constructs
neutron stars using Oppenheimer -- Volkoff equations and
appropriate equations of state. In the case of neutron stars,
one has a very good knowledge about the nature and the
properties of the constituents, namely protons, neutrons,
quarks, gluons, et cetera. In the case of horizonless objects,
which can be macroscopic, such knowledge is absent at present.
These horizonless objects may be the massive remnants suggested
in \cite {sbg} whose size and mass depend on the information
contained; or they may be the ones which develop `firewalls' as
suggested in \cite{amps, amps2, amps3}; or they may be the `fuzz
balls' of Mathur's proposal which have no horizon at anytime
\cite {fuzz, fuzz2, fuzz3}. Recently, we have also argued for
the existence of such horizonless fuzz ball like objects in any
quantum theory of gravity where singularities are resolved and
evolutions are unitary \cite {k12, k122, k123}. A variety of
horizonless objects have been advocated in the past also from
several points of view, see \cite{kg} -- \cite {page12} for a
sample of them.

Although nothing is known rigorously about the horizonless
objects and their constituents, it is physically reasonable to
expect them to have the following properties. {\bf (i)} Their
central densities are likely to be of the order of Planckian
densities since the quantum gravity effects are expected to play
an important role in creating them. {\bf (ii)} Their sizes are
likely to be of the order of Schwarzschild radii since they are
expected to be similar to black holes, only without horizons.
{\bf (iii)} Being the end products of prior gravitational
collapses and evolutions, they must be stable against any
further gravitational collapse. {\bf (iv)} Their constituents
must have a large number of entropic degrees of freedom and,
hence, thermodynamics must be applicable. Therefore these
constituents, even if highly quantum in nature, may be modelled
using density, pressures, and suitable equations of state. {\bf
(v)} These horizonless objects can be arbitrarily massive.
Hence, atleast qualitatively, the corresponding spacetime may be
described by general relativity equations with appropriate
energy momentum tensors and constituent equations of state.

In this paper, we consider static, spherically symmetric cases
and focus on property {\bf (iii)}, namely the stability
property. In the following, for the sake of brevity, we will
refer to these horizonless objects as stars although they are
not stars in a conventional sense. For example, their central
densities are likely to be of the order of Planckian densities;
and their sizes are likely to be of the order of Schwarzschild
radii. 

Sufficiently massive conventional stars have been found to be
unstable against gravitational collapse. But, in his study of
relativistic stars in arbitrary dimensions \cite{pc07}, see also
\cite{pc072}, Chavanis found among other things that stars can
be stable against collapse, albeit only in higher dimensions. To
the best of our knowledge, this is the only work which has shown
the possibility of arbitrarily massive stars being also stable;
moreover, the radii of these stable stars scale as Schwarzschild
radii. \footnote{ Similar results were also obtained around the
same time in \cite {ads11, ads112}, but in asymptotically anti
de Sitter spacetimes whose radii limit the masses of the stable
stars.}  Chavanis considered static spherically symmetric stars
made up of a perfect fluid with a linear equation of state and
enclosed them in a box. \footnote{ This box is needed since the
radius of the star will be infinite otherwise \cite{rs}. As
explained in \cite{pc07}, enclosing the star within a box
prevents the evaporation of its constituents and makes its total
mass finite. The radius and the mass may also be made finite by
constructing a composite configuration consisting of a perfect
fluid core and a crust of constant density or a gaseous envelope
exerting a constant pressure, see \cite {cpaper, cpaper2,
cpaper3}. \label{box}} Following the methods of Chandrasekhar
\cite{cpaper}, he then obtained `singular solutions' and
analysed the asymptotic perturbations around them. He has shown
in detail that the damped oscillatory, or monotnic non
oscillatory, behaviour of these perturbations around singular
solutions leads to the damped oscillatory, or monotnic non
oscillatory, behaviour for the mass -- central density profile
of the star in the asymptotic region of large central density or
large radius. He has shown further that equilibrium
configurations become unstable beyond the first maximum in this
profile, hence its oscillatory behaviour in the asymptotic
regions will imply instability against collapse. Studying the
effects of higher dimensions, he then found that the mass --
central density profiles of the stars in eleven or higher
dimensions are non oscillatory in both the asymptotic and non
asymptotic regions, and imply stability.

If the number eleven above counts both compact and non compact
directions then stability in, for example, four dimensional non
compact spacetime may be obtained by having a eleven dimensional
spacetime with seven compact dimensions, which is quite natural
in the context of M theory. With this motivation, in \cite{k13},
we generalised the work of Chavanis to include compact
directions and multi component perfect fluids, the later also
appearing quite naturally in string and M theory black branes.
However, we found that, even in these generalised cases, eleven
non compact dimensions are needed for stability. In particular,
this implied that stars in four dimensions and those in M theory
are unstable and that they will collapse if sufficiently
massive.

Reviewing the underlying assumptions in \cite{pc07, k13} and
studying carefully some of the works on the horizonless objects,
particularly \cite{cl} -- \cite{cfv}, we realised that the 
pressures inside the stars along the noncompact spatial
directions need not be isotropic. The assumption about isotropy
may be unwarranted and restrictive : it is not required by
spherical symmetry and, in general, the pressures may be
different along the radial and the transverse spherical
directions of the non compact space. For conventional boson
stars, anisotropic pressures arise naturally when scalar fields
are present \cite {mg, mg2}. See also the reviews \cite{mg3,
mg4, mg5, mg6} and the recent paper \cite {bnv}. For the
horizonless objects, which are implied by unitarity in lieu of
black holes but are referred to as stars here, there may be
other mechanisms giving rise to anisotropic pressures. However,
since not much is known about the constituents of such objects,
we do not know any possible exact mechanisms.

In this paper, therefore, we simply assume that pressures in the
stars may be anisotropic and study, a la Chavanis, the effects
of such anisotropy on the stability of stars. Following closely
the works in \cite{pc07, cpaper, k13} and, generalising them now
by including anisotropic cases, we study stars in higher
dimensional spacetime which may have compact toroidal directions
also. The stars are assumed to be static and spherically
symmetric in the non compact space and to have suitable
isometries along the compact directions. Taking a suitable
ansatz for the metric, we write down the equations of
motion. Then, as in \cite{pc07, cpaper}, we assume linear
equations of state and obtain the `singular solutions' and the
asymptotic perturbations around them. The oscillatory or non
oscillatory behaviour of these perturbations lead to
corresponding behaviour for the mass -- central density profile
in the asymptotic regions. We assume that, as in Chavanis' works
\cite{pc07}, the equilibrium configurations become unstable
beyond the first maximum in the mass -- central density profile;
hence that the damped oscillatory behaviour of this profile
implies instability whereas monotnic non oscillatory behaviour
throughout implies stability. Studying then the asymptotic
perturbations around the `singular solutions', we obtain the
criteria under which the perturbations are non oscillatory.
These will then be the necessary criteria for stability.
\footnote{ \label{nonasymp} One must now show that the
equilibrium configurations in the mass -- central density
profile become unstable beyond the first maximum. And, for the
case where asymptotic behaviour is non oscillatory, one must
show that this profile remains monotonic everywhere, including
the non asymptotic region also. Showing these is beyond the
scope of the present paper since it requires a knowldege of the
nature and the properties of constituents of the horizonless
objects, which is lacking at present.} We perform the above
analysis first for stars in spacetime with no compact
directions; then include compact directions; then repeat the
analysis for two examples of stars in M theory made up of stacks
of (intersecting) two branes and five branes. The formulation
presented in this paper may, however, be used to study a variety
of other examples also.

We find that, in an $(m + 2)$ dimensional non compact spacetime
with $m \ge 2 \;$, non oscillatory perturbations are possible
for any value of $m$ if an appropriate amount of anisotropy is
present, namely if
\[
\frac {\Pi - p} {\Pi} \; \ge \; \frac {9 - m} {4 m}
\]
where $\Pi$ and $p$ are the pressures along the radial and the
transverse spherical directions. Also, we find that presence of
compact directions does not change this result. Similar result
follows for stars in M theory also. It can now be seen that if
$m \ge 9$, namely if the non compact spacetime is eleven or
higher dimensional, then the isotropic case (above ratio $= 0$)
is included in the above range, thus reproducing the result in
\cite{pc07}. For lower values of $m$, certain amount of
anisotropy is needed to obtain non oscillatory perturbations;
for example, for four dimensional spacetime, $m = 2$ and the
above ratio needs to be $ \ge \frac {7} {8} \;$. 

This result suggests that it may be possible to have stable
horizonless objects in four or any higher dimensions, and that
anisotropic pressures may play a crucial role in ensuring their
stability. Although much remains to be done, for example resolve
the issues mentioned in footnote {\bf \ref{nonasymp}}, it is
worth emphasising that this is an important result because it
bears on the horizonless objects which are implied by unitarity
in lieu of black holes, and it points out a necessary ingredient
for their stability. To actually construct such objects,
however, requires detailed understanding of many issues such as
the nature of the constituents and the physical mechanisms that
may provide the required amount of anisotropy. We will discuss
briefly these and other issues at the end of the paper.

This paper is organised as follows. In section {\bf 2}, we first
mention briefly the relevant aspects of \cite{pc07} which are
used here. We then briefly describe M theory stars and mention
how the equations of state are obtained. In section {\bf 3}, we
set up our notations and conventions and present the equations
of motion in a suitable form. In section {\bf 4}, we give the
details of the asymptotic analysis, which involve singular
solutions and perturbations around them, and outline briefly the
significance of perturbations. In section {\bf 5}, we obtain
singular solutions for stars in spacetime with no compact
directions, study the perturbations around them, and obtain the
criteria for their non oscillatory behaviour. In section {\bf
6}, we repeat the analysis for two examples of stars in M theory
: in one, stars are made up of a stack of two branes or five
branes; in another, stars are made up of four stacks of
intersecting branes, two stacks each of two branes and five
branes. In section {\bf 7}, we conclude with a brief summary and
a discussion of some of the issues which require further study.


\vspace{4ex}

\centerline{\bf 2. Some general remarks} 

\vspace{2ex}

In this section, we first mention briefly some aspects of
\cite{pc07} which are relevant here. For more details, see \cite
{pc07} and the references there. We then briefly describe M
theory stars and mention how the equations of state are
obtained.

Chavanis considers static spherically symmetric star with its
constituents obeying linear equations of state. In cosmological
contexts, it is a standard practice to use linear equations
state. In the contexts of stars, Chavanis explains several
situations where such equations of state arise and the reasons
for using them. The star is then enclosed in a box of radius
$r_*$ which prevents the evaporation of its constituents and
makes its total mass finite. The general relativistic equations
are then solved to obtain hydrostatic equilibrium
configurations.

Considering a series of equilibria, the mass -- central density
profile is obtained and shown to have damped oscillatory
behaviour. The hydrostatic equilibrium configurations become
unstable beyond the first maximum in this profile. This is also
shown to correspond to conditions for nonlinear dynamic
stability. Thus, for a given volume, the star has a maximum mass
above which it becomes unstable.

Following Chandrasekhar \cite {cpaper}, singular solutions and
the asymptotic perturbations around them are obtained. The
singular solutions give the scaling relations between mass,
radius, and other quantities of the stars. The damped
oscillatory, or monotonic non oscillatory, behaviour of the
perturbations are shown to be responsible for the corresponding
behaviour for the mass -- central density profile in the
asymptotic region. 

Chavanis obtains the mass -- central density profile in the non
asymptotic regions also. He shows that the profile in $D <
D_{crit}$ dimensional spacetime starts from origin, increases
monotonically, reaches a maximum, then decreases, and oscillates
with damped amplitudes around the constant value line given by
the singular solutions; whereas, in $D \ge D_{crit}$ dimensional
spacetime, the entire profile is monotonically increasing, and
asymptotes to the constant value line given by the singular
solutions.  The critical dimension $D_{crit} \simeq 11$ and
depends on the pressure to density ratio. See Figure 23 in
\cite{pc07}.

If the mass -- central density profile is oscillatory then the
scaling behaviour given by the singular solutions is not stable
since the stability is lost beyond the first maximum of the
profile. The singular solutions are then of no relevance. The
mass -- radius relations and other quantities of stable stars
must be obtained from detailed analysis of the equations in the
non asymptotic regions before the first maximum.

If the entire mass -- central density profile is monotonic and
non oscillatory then the series of equilibria represented by its
points are all stable. Then the singular solutions correspond to
stable configurations and they can be used to obtain 
mass -- radius relations and other quantities of stable stars
in the limit of large central density or large radius. 

In the case of horizonless objects which are implied by
unitarity and which we study here, nothing is known rigorously
about the nature and properties of their constituents. Hence a
thorough analysis as in \cite{pc07} is presently not possible.
Nevertheless, it is possible to make some progress using the
physically reasonable properties of such objects listed in the
Introduction. We assume a linear equation of state for the
constituents. The stiffest equation of state ($pressure =
density$, $\; sound \; speed = light \; speed$) is of this type.
Also, this linearity helps in obtaining analytically the
singular solutions and the asymptotic perturbations around them,
and in studying whether monotonic non oscillatory behaviour is
possible. We enclose the star in a box of radius $r_*$ which
prevents the evaporation of its constituents and makes its total
mass finite. Then, as in \cite {pc07}, the damped oscillatory,
or monotonic non oscillatory, behaviour of these perturbations
can be shown to lead to the corresponding behaviour for the mass
-- central density profile of the star in the asymptotic
region. However, we can not obtain this profile in the non
asymptotic regions, nor study the stability properties around
the maxima, without knowing the detailed properties of the
constituents.

Detailed knowledge of the constituents of the horizonless
objects is also needed in order to understand the fate of, for
example, a collapsing massive neutron star which would have
formed a black hole in the standard scenario. This knowledge is
needed to understand what happens to neutrons, protons, et
cetera in the quantum gravity regime where the singularities are
assumed to be resolved. This requires understanding the relation
between quantum gravity theory and standard model particles.


\vspace{2ex}

\centerline{\bf M theory stars}

\vspace{2ex}

In String/M theory, the entropy and Hawking radiation of a class
of extremal and near extremal black holes have been understood
rigorously in terms of various intersecting brane configurations
and the low energy excitations on them. Mathur's fuzz ball
proposal also arises naturally in this context. The entropy and
Hawking radiation of neutral or far from extremal black holes
are not understood equally rigorously, but their explanations
are likely to be in terms of intersecting brane antibrane
configurations along the lines given in \cite {hm} -- \cite
{k046}.
In
the intersecting brane configurations, various stacks of branes
wrap around compact toroidal directions, intersecting according
to BPS rules whereby, in M theory, \footnote{ The brane
configurations in string and M theories are equivalent and are
related by chains of U duality operations involving dimensional
reduction and upliftment between string and M theory, and the T
and S dualities of the string theories. Hence, in the following,
we will restrict ourselves to M theory. The corresponding string
theory results are straightforward to obtain.} two stacks of
five branes intersect along three common spatial directions; a
stack each of two branes and five branes intersect along one
common spatial direction; and two stacks of two branes intersect
along zero common spatial direction. See \cite{alg} and the
references therein. Given the understanding of near extremal
black hole properties in terms of intersecting branes, and given
the fuzz ball proposal, it is natural to expect that there must
be stars in M theory made up of intersecting branes \footnote{ M
theory stars which are considered here and are made up of, for
example, two branes may be visualised as follows. The spacetime
is eleven dimensional with two compact toroidal spatial
directions. The non compact space is eight dimensional,
described by polar coordinates $(r, \; \theta_1, \; \cdots, \;
\theta_7) \;$. Thus there is a torus at every point of the non
compact space. Stacks of two branes wrap around these tori such
that there are a total of $n(r)$ number of two branes per unit
volume of the non compact space. This number density $n(r) \ne
0$ for $r < r_* \;$ and $= 0$ for $r \ge r_* \;$. This provides
a picture of an M theory star of radius $r_*$ made up of two
branes. This picture is analogous to that of a four dimensional
star of radius $r_*$ where $(r, \; \theta_1, \theta_2)$ are the
polar coordinates and $n(r)$ is the particle density. The stack
of branes wrapping the torus at a point is like a `particle' in
the non compact space. One may similarly visualise M theory
stars made up of other intersecting brane configurations. A
`particle' in the non compact space now is the ${\cal N}$ stacks
of two and five branes wrapping the tori at a point, with
necessary isometries, and intersecting according to BPS
rules. This is a rough picture; the details of how the
`particles' interact are far from being rigorously known.
\label{ftmstar} } which, when sufficiently massive, will
collapse and form either black holes having horizon or fuzz ball
like objects having no horizon. Also, for the same entropic
reasons which are explained in \cite{cm, cm2, k07} where early
universe was studied in string/M theory, we may assume that
stars in M theory are made up of stacks of intersecting branes.

M theory stars may be studied using the formalism given in this
paper. Assuming the spatial directions of the brane worldvolumes
to be toroidal and assuming necessary isometries, the M theory
brane configurations consisting of ${\cal N}$ stacks of two and
five branes which intersect according to BPS rules, can be
described by ${\cal N}$ seperately conserved energy momentum
tensors $T_{M N (I)}$, $\; I = 1, 2, \cdots, {\cal N} \;$, with
appropriate equations of state among their components. See
\cite{aeh, aeh2, aeh3} and the references therein. The equations
of state may follow from an action; or they may be derived using
the microscopic dynamics of the constituents which are far from
being rigorously known; or they may simply be postulated as an
ansatz. \footnote{ This situation is similar to that for, for
example, charged black holes versus for stars made up of charged
particles or for an expanding universe containing them: In the
case of black holes, equations are obtained from an action. In
the case of stars, or an expanding universe, one may use
statistical mechanics to describe charges, anticharges, and the
photons between them; or one may simply take, for example, a
linear equation of state as an ansatz.}

In M theory, equations describing black hole spacetimes follow
from an eleven dimensional low energy effective action \cite
{hm, hm2, alg, aeh, aeh2, aeh3}. In the context of an expanding
universe, the equations of state for intersecting branes have
been derived in certain approximations using microscopic
dynamics of branes \cite {cm, cm2}. In the context of an
expanding universe, and also of stars, we used U duality
symmetries and derived a relation among the components of the
energy momentum tensor \cite{k07}.  \footnote{ Using dimensional
reduction and upliftment, T dualities, and S dualities, we
obtained the U duality relations first in the cosmological
context in \cite {k07} and then, in unpublished notes, for stars
and black branes also. See the comments and the application of
these U duality relations in \cite{k07U}. \label{ftk07U}} These
relations, one each for each of the ${\cal N}$ stacks in the
intersecting brane configurations, follow as a consequence of U
duality symmetries and, therefore, must always be valid
independent of the details of the equations of state. The
equations of state in \cite {hm, hm2, alg, aeh, aeh2, aeh3, cm,
cm2}, which were obtained by other methods, all obey these U
duality relations.

The U duality relations need to be supplemented with the
equations of state in, say $d$ dimensional, non compact part of
the total eleven dimensional spacetime. These data on the
equations of state are same as those needed for any expanding
universes or for any stars in $d$ dimensional non compact
spacetimes but now with no compact dimensions. See \cite {k07,
bdr, bdr2, bdr3, k13} for examples and more details, as well as
for applications and non trivial consequences of these U duality
relations for expanding universes and for stars in M theory.

Note that, in the context of M theory also, understanding the
fate of, for example, a collapsing massive neutron star which
would have formed a black hole in the standard scenario,
requires understanding the relation between M theory and
standard model particles.


\vspace{4ex}

\centerline{\bf 3. Equations of motion}

\vspace{2ex}

In this paper, we consider static cases which are spherically
symmetric in higher dimensional spacetime. We will also consider
such stars in eleven dimensional M theory. Although the
formalism and the results in this paper are applicable to
conventional stars also, our main interest here is in the
horizonless objects and their stability properties. As explained
in the Introduction, these objects are implied by unitarity in
lieu of black holes and, for the sake of brevity, are also
referred to as stars here. They are not stars in a conventional
sense because, for example, their central densities are likely
to be of the order of Planckian densities; and their sizes are
likely to be of the order of Schwarzschild radii. 

The spacetime is assumed to be $D = n_c + m + 2$ dimensional
with $m \ge 2 \;$, with $n_c$ dimensional compact toroidal
space, and with $(m + 1)$ dimensional non compact space. The
stars are assumed to be static and spherically symmetric in the
non compact space, and to be made up of non interacting
multicomponent fluids with linear equations of state. The M
theory stars are eleven dimensional and are assumed to be made
up of ${\cal N}$ stacks of $M 2$ and $M 5$ branes, intersecting
according to the BPS rules whereby two stacks of five branes
intersect along three common spatial directions; a stack each of
two branes and five branes intersect along one common spatial
direction; and two stacks of two branes intersect along zero
common spatial direction. Assuming the spatial directions of the
brane worldvolumes to be toroidal and assuming necessary
isometries, the intersecting M theory branes can be described by
${\cal N}$ seperately conserved energy momentum tensors $T_{M N
(I)}$, $\; I = 1, 2, \cdots, {\cal N} \;$, with appropriate
equations of state among their components \cite{alg, aeh, aeh2,
aeh3, cm, cm2, k07, bdr, bdr2, bdr3}.

Following closely the notations and conventions of our earlier
work \cite{k13}, we now write a suitable ansatz for the metric
and obtain the equations of motion. Let $x^M = (t, x^i, r,
\theta^a)$ be the spacetime coordinates where $x^i$, $\; i = 1,
2, \cdots, n_c \;$, describe the $n_c$ dimensional toroidal
space; and the radial and the spherical coordinates $(r, \;
\theta^a)$, $\; a = 1, 2, \cdots, m \;$, describe the $(m + 1)$
dimensional non compact space. In standard notation and with
$\kappa^2 = 8 \pi G_D = 1 \;$, the equations of motion may be
written as
\begin{eqnarray}
{\cal R}_{M N} - \frac{1}{2} \; g_{M N}\; {\cal R} & = &
T_{M N} \; = \; \sum_I T_{M N (I)} \label{rmn} \\
\sum_M \; \nabla_M \; T^M_{\; \; N (I)} & = & 0  \label{tmnI}
\end{eqnarray}
where $T_{M N}$ is the total energy momentum tensor of non
interacting multicomponent fluids and $T_{M N (I)}$ is the
energy momentum tensor for the $I^{th}$ component fluid. For
stars in M theory, $T_{M N}$ is the total energy momentum tensor
for intersecting branes and $T_{M N (I)}$ is the energy momentum
tensor for the $I^{th}$ stack of branes. 

In the following, we consider static solutions which are
spherically symmetric in the $(m + 1)$ dimensional non compact
space. We will study the singular solutions and the asymptotic
perturbations around them in the limit of large $r$ \cite
{cpaper, pc07}. The suitable ansatz for the line element $d s$
is given by
\begin{equation}\label{ds} 
d s^2 = g_{M N} \; d x^M \; d x^N 
= - e^{2 \lambda^0} d t^2 + \sum_i e^{2 \lambda^i} (d x^i)^2
+ e^{2 \lambda} d r^2 + e^{2 \sigma} d \Omega_m^2
\end{equation}
where $d \Omega_m$ is the standard line element on an $m$
dimensional unit sphere. The energy momentum tensors $T_{M N
(I)}$ are assumed to be diagonal. These diagonal elements are
denoted as
\[
\left( T^0_{\; \; 0 (I)}, \; T^i_{\; \; i (I)}, \; 
T^r_{\; \; r (I)}, \; T^a_{\; \; a (I)} \right) = 
\left( p_{0 I}, \; p_{i I}, \; \Pi_I, \; p_{a I} \right) 
\]
where $p_{0 I} = - \rho_I$ and $p_{a I} = p_I$ for all $a \;$.
The total energy momentum tensor is now given by $T^M_{\; \; \;
\; N} = diag \; \left( p_0, \; p_i, \; \Pi, \; p_a \right)$
where $p_0 = - \rho \;$, $\; p_a = p$ for all $a \;$, and
\[
\rho = \sum_I \rho_I \; \; , \; \; \; 
p_i = \sum_I p_{i I} \; \; , \; \; \; 
\Pi = \sum_I \Pi_I \; \; , \; \; \; 
p = \sum_I p_I \; \; .
\]
Define
\[
\alpha = (0, i, a) 
\; \; , \; \; \; 
\lambda^\alpha = (\lambda^0, \lambda^i, \lambda^a) 
\; \; , \; \; \; 
p_{\alpha I} = (p_{0 I}, p_{i I}, p_{a I}) 
\]
where $\lambda^a = \sigma$ for all $a \;$. Also, define
\[
\Lambda = \sum_\alpha \lambda^\alpha = 
\lambda^0 + \sum_i \lambda^i + m \sigma 
\; \; , \; \; \; 
T_I = \sum_M T^M_{\; \; \; M (I)} = 
\Pi_I + \sum_\alpha p_{\alpha I} \; \; . 
\]
For static solutions which are spherically symmetric in the 
$(m + 1)$ dimensional non compact space, the fields
$(\lambda^\alpha, \; \lambda)$ and $(p_{\alpha I}, \Pi_I)$
depend only on the coordinate $r \;$. Using the above
definitions and the metric given in equation (\ref{ds}), it
follows straightforwardly that the equations of motion
(\ref{rmn}) and (\ref{tmnI}) now give
\begin{eqnarray}
(\Pi_I)_r & = & - \Pi_I \; \Lambda_r 
+ \sum_\alpha p_{\alpha I} \lambda^\alpha_r \label{Pir} \\
\Lambda^2_r - \sum_\alpha (\lambda^\alpha_r)^2 & = &
2 \sum_I \Pi_I \; e^{2 \lambda} 
+ m (m - 1) \; e^{2 \lambda - 2 \sigma} \label{Lambdar2}
\end{eqnarray}
\begin{equation}\label{alpharr}
\lambda^\alpha_{r r} + (\Lambda_r - \lambda_r) \;
\lambda^\alpha_r \; = \; \sum_I \left(- p_{\alpha I} 
+ \frac{T_I}{D - 2} \right) \; e^{2 \lambda} 
+ \delta^{\alpha a} \; (m - 1) \; e^{2 \lambda - 2 \sigma}
\end{equation}
where the subscripts $r$ denote $r-$derivatives. We also define
a function $f(r)$ and a mass function $M(r)$ by
\[
e^{2 \lambda - 2 \sigma} = \frac{1}{r^2 f} 
\; \; \; , \; \; \; \; \; 
f = 1 - \frac{M}{r^{m - 1}} 
\]
so that either of them may be traded for the function $\lambda
(r) \;$.


\vspace{2ex}

\centerline{\bf Reduction to $d = m + 2$ dimensions} 

\vspace{2ex}

We will now dimensionally reduce on the $n_c$ dimensional
toroidal space from $D$ dimensions to $d = m + 2$ dimensions
described by the $x^\mu = (t, r, \theta^a)$ coordinates.
Consider the $D$ dimensional line element $d s$ given by
equation (\ref{ds}), and denote its $d$ dimensional part as
follows:
\[
d s^2_d = g_{\mu \nu (d)} \; d x^\mu \; d x^\nu = 
- e^{2 \lambda^0} d t^2 + e^{2 \lambda} d r^2 
+ e^{2 \sigma} d \Omega_m^2 \; \; . 
\]
Upon dimensional reduction, symbolically, we have 
\begin{eqnarray*}
S & \sim & \int d^D x \; \sqrt{- g} \; R \; \sim \; \int d^d x
\; \sqrt{- g_{(d)}} \; e^{\Lambda^c} \; (R_{(d)} + \cdots) \\
& & \\
& \sim & \int d^d x \; \sqrt{- \tilde{g}} \; 
(\tilde{R} + \cdots)
\end{eqnarray*}
where $\Lambda^c = \sum_i \lambda^i$ and $\tilde{g}_{\mu \nu} =
e^{\frac{2 \Lambda^c}{m}} \; g_{\mu \nu (d)} \;$ is the $d$
dimensional Einstein frame metric. The corresponding line
element $\tilde{d s}_d$ then becomes
\begin{equation}\label{dsd}
\tilde{d s}^2_d = \tilde{g}_{\mu \nu} \; d x^\mu \; d x^\nu = 
- e^{2 \tilde{\lambda}^0} d t^2 + e^{2 \tilde{\lambda}} d r^2 
+ e^{2 \tilde{\sigma}} d \Omega_m^2 
\end{equation}
where 
\begin{equation}\label{ltil1}
\tilde{\lambda}^\alpha = \lambda^\alpha + \frac{\Lambda^c}{m}
\; \; \; , \; \; \; \; 
\tilde{\lambda} = \lambda + \frac{\Lambda^c}{m} \; \; , 
\end{equation}
and $\tilde{\lambda}^a = \tilde{\sigma}$ for all $a \;$.
Furthermore, one has
\[
\tilde{\Lambda} = \tilde{\lambda}^0 + m \tilde{\sigma} 
= \Lambda + \frac{\Lambda^c}{m} \; \; , \; \; \;
e^{2 \tilde{\lambda} - 2 \tilde{\sigma}} = 
e^{2 \lambda - 2 \sigma} = \frac{1}{r^2 f} \; \; .
\]
Note that $\tilde {\lambda}^i$ and $\lambda^i \;$ can be
expressed easily in terms of each other \footnote{ For any
$a^\alpha$s, let $\tilde {a}^\alpha = a^\alpha + \frac{a^c} {m}$
where $a^c = \sum_i a^i \;$. Then $\tilde{a}^c = \sum_i \tilde
{a}^i = (n_c + m) \frac{a^c} {m}$ and, hence, $a^\alpha = \tilde
{a}^\alpha - \frac {\tilde {a}^c} {n_c + m} \;$. Similarly for
any $b^\alpha$s. Also, $\sum_i \tilde {a}^i b^i = \sum_i a^i
\tilde {b}^i = \sum_i a^i b^i + \frac {a^c b^c} {m}$, which
further implies that $\sum_i \tilde {a}^i a^i > 0$ if $a^i$ do
not all vanish. \label{untilde}} and, hence, they are both
equally convenient to work with.

Writing $p_{0 I} = - \rho_I$, it now follows from equations
(\ref{Pir}) -- (\ref{alpharr}) that
\begin{equation}\label{tldPir}
(\Pi_I)_r = - \Pi_I \; \tilde{\Lambda}_r 
- \rho_I \; \tilde{\lambda}^0_r 
+ m \; p_I \; \tilde{\sigma}_r 
+ \sum_i \left( p_{i I} - \frac{{\cal T}_I}{m} \right)
\lambda^i_r 
+ \frac {2 \Pi_I} {m} \; \Lambda^c_r 
\end{equation}
\begin{eqnarray}
2 \tilde{\lambda}^0_r \tilde{\sigma}_r 
+ (m - 1) (\tilde{\sigma}_r)^2 & = &
\frac{2}{m} \sum_I \Pi_I \; e^{2 \lambda} 
+ (m - 1) \; e^{2 \tilde{\lambda} - 2 \tilde{\sigma}} 
+ \frac{{\cal B}}{m} \label{tldLambdar2} \\
\tilde{\sigma}_{r r} + (\tilde{\Lambda}_r - \tilde{\lambda}_r)
\; \tilde{\sigma}_r & = & \sum_I \left(- p_I
+ \frac{{\cal T}_I}{m} \right) \; e^{2 \lambda} + (m - 1) \;
e^{2 \tilde{\lambda} - 2 \tilde{\sigma}} \label{tldsigmarr} \\
\tilde{\lambda}^0_{r r} + (\tilde{\Lambda}_r 
- \tilde{\lambda}_r) \; \tilde{\lambda}^0_r & = & 
\sum_I \left(\rho_I + \frac{{\cal T}_I}{m} \right) \; 
e^{2 \lambda} \label{tld0rr} \\
\tilde{\lambda}^i_{r r} + (\tilde{\Lambda}_r 
- \tilde{\lambda}_r) \; \tilde{\lambda}^i_r & = & 
\sum_I \left(- p_{i I} + \frac{{\cal T}_I}{m} \right) \; 
e^{2 \lambda} \label{tldirr} 
\end{eqnarray} 
where ${\cal T}_I = \Pi_I - \rho_I + m \; p_I$ and
${\cal B} = \sum_i \tilde{\lambda}^i_r \; \lambda^i_r \; $.
Using the diffeomorphic freedom in defining the radial
coordinate, we now set $e^{\tilde{\sigma}} = r \;$. Equations
(\ref{tldLambdar2}) and (\ref{tldsigmarr}) become

\begin{eqnarray}
r \; \tilde{\lambda}^0_r & = & \sum_I \frac{\Pi_I}{m} \; r^2
\; e^{2 \lambda} + \frac{m - 1}{2} \; ( 
e^{2 \tilde{\lambda}} - 1 ) + \frac{r^2 {\cal B}}{2 m} 
\label{2tldLambdar2} \\
& & \nonumber \\
r \; (\tilde{\lambda}^0_r - \tilde{\lambda}_r) & = & \sum_I
\left(\frac {\Pi_I - \rho_I} {m} \right) \; r^2 \; e^{2 \lambda}
+ (m - 1) \; (e^{2 \tilde{\lambda}} - 1 ) \label{2tldsigmarr} \\
& & \nonumber \\
\Longrightarrow \; \; \; \; 
r \; \tilde{\lambda}_r & = & \sum_I \frac{\rho_I}{m} \; r^2 \;
e^{2 \lambda} - \frac{m - 1}{2} \; ( e^{2 \tilde{\lambda}} - 1 )
+ \frac{r^2 {\cal B}}{2 m} \; \; . \label{2tldLambdar22} 
\end{eqnarray}
Since $e^{\tilde{\sigma}} = r \;$, we also have
\[
\tilde{d s}^2_d = - e^{2 \tilde{\lambda}^0} d t^2 
+ \frac{d r^2}{f} + r^2 d \Omega_m^2
\; \; \; , \; \; \; 
f = e^{- 2 \tilde{\lambda}} = 1 - \frac{M}{r^{m - 1}} \; \; .
\] 
Thus, the line element $\tilde{d s}_d$ in the $d = m + 2$
dimensional Einstein frame takes the standard form. Therefore
the functions appearing in it, {\em e.g.} the mass function $M$,
may be interpreted in the standard way. For instance, $M_{ADM}$,
the ADM mass of the star of radius $r_*$ is related to the mass
function by $M(r_*) = \frac {16 \; \pi \; G_D} {m \; S_m \;
V_{n_c}} \; M_{ADM}$ where $S_m = \frac {2 \; \pi^{ \frac{m +
1}{2} }} {\Gamma (\frac{m + 1}{2})}$ is the `area' of an $m$
dimensional unit sphere and $V_{n_c}$ is the coordinate volume
of the $n_c$ dimensional toroidal space. Note that equation
(\ref{2tldLambdar22}) and the relation $M(r) = r^{m - 1} \; ( 1
- e^{- 2 \tilde{\lambda}} ) \;$ now give
\begin{equation}\label{mr}
M_r \; = \; \frac{r^m}{m} \; \left( 2 \sum_I \rho_I \; 
e^{- \frac{2 \Lambda^c}{m}} + {\cal B} \; 
e^{- 2 \tilde{\lambda}} \right) \; \; . 
\end{equation} 


\vspace{2ex}

\centerline{\bf Linear equations of state}

\vspace{2ex}

To solve equations (\ref{tldPir}) and (\ref{tld0rr}) --
(\ref{2tldsigmarr}), and to obtain solutions for the fields
$(\lambda^\alpha, \; p_{\alpha I}, \; \Pi_I)$, one further
requires equations of state which give $p_{\alpha I}$ and
$\Pi_I$ as functions of $\rho_I \;$. In cosmological contexts,
it is a standard practice to use linear equations state. In the
contexts of stars, Chavanis explains several situations where
such equations of state arise and the reasons for using them
\cite{pc07}.  In the case of horizonless objects which are
implied by unitarity and which are of main interest here,
nothing is known rigorously about the nature and properties of
their constituents. Hence, it is not presently possible to
derive the equations of state from the underlying microscopic
physics. In order to make progress, we will assume here linear
equations of state. The stiffest equation of state ($pressure =
density$, $\; sound \; speed = light \; speed$) is of this type.
Also, the linearity helps in obtaining explicitly the singular
solutions and the asymptotic perturbations around them, and in
studying the stability properties.

Thus, let
\begin{equation}\label{eoswi}
p_{\alpha I} = w^I_\alpha \; \rho_I \; \; , \; \; \; 
\Pi_I = w^I_\pi \; \rho_I
\end{equation}
where $w^I_\alpha$ and $w^I_\pi$ are constants, $w^I_0 = - 1$
since $p_{0 I} = - \rho_I$, $w^I_a = w^I$ since $p_{a I} = p_I$
for all $a \;$, and we assume that $w^I_\pi > 0 \;$. It is
common to take the pressures inside the stars to be isotropic
along the noncompact spatial directions, namely to take the
pressure $\Pi_I$ along the radial direction to be equal to the
pressure $p_I \;$ along the transverse spherical directions.
However, the assumption about isotropy may be unwarranted and
restrictive : it is not required by spherical symmetry and, in
general, the pressures inside the stars may be different along
the radial and the transverse spherical directions of the non
compact space. For conventional boson stars, anisotropic
pressures arise naturally when scalar fields are present \cite
{mg, mg2}. See also the reviews \cite{mg3, mg4, mg5, mg6} and
the recent paper \cite {bnv}. We will assume here that such an
anisotropy may be present and, hence, that $\Pi_I \ne p_I \;$ in
general. As a measure of this anisotropy, we define a
dimensionless parameter $\eta^I$ by
\begin{equation}\label{etai} 
\eta^I = \frac {m} {2} \; \left( \frac {\Pi_I - p_I} {\Pi_I}
\right) = \frac {m} {2} \; \left( \frac {w^I_\pi - w^I}
{w^I_\pi} \right)
\end{equation} 
so that $\eta^I = 0$ corresponds to the isotropic case. The
factor of $\frac {m} {2}$ is for convenience.

Since ${\cal T}_I = \Pi_I - \rho_I + m \; p_I$, equations
(\ref{eoswi}) give
\[
\left( - p_{\alpha I} + \frac{{\cal T}_I}{m} \right) =
\tilde{c}^{\alpha I} \; \rho_I \; \; , \; \; \; \;
\tilde{c}^{\alpha I} = - w^I_\alpha + w^I 
+ \frac {w^I_\pi - 1} {m} \; \; .
\]
Hence
\begin{equation}\label{tldcalpha} 
\tilde{c}^{0 I} = 1 + w^I + \tilde{c}^I \; \; , \; \; \;
\tilde{c}^{i I} = - w^I_i + w^I + \tilde{c}^I 
\; \; , \; \; \;
\tilde{c}^{a I} = 
\tilde{c}^I = \frac {w^I_\pi - 1} {m} \; \; . 
\end{equation}
The corresponding untilded coefficients are given by $c^{\alpha
I} = \tilde{c}^{\alpha I} - \frac {\sum_j \tilde {c}^{j I} }
{n_c + m} \;$, see footnote {\bf \ref{untilde}}. Let $\phi^I$ be
given by
\begin{equation}\label{phii} 
w^I_\pi \; \phi^I \; = \; - \; (1 + w^I_\pi) \;
\tilde{\lambda}^0 - 2 \; \eta^I w^I_\pi \; \tilde{\sigma} 
- \sum_i c^{i I} \; \tilde{\lambda}^i \; \; . 
\end{equation}
It then follows from equations (\ref{tldPir}) and (\ref{etai}),
and from $e^{\tilde {\sigma}} = r$, that
\begin{equation}\label{rho}
\rho_I \; = \; \rho_{I 0} \; e^{\phi^I} \; 
e^{\frac {2 \Lambda^c} {m}}
\; \; \; , \; \; \; \; 
r^2 \rho_I \; e^{2 \lambda} = \rho_{I 0} \; e^{\phi^I 
+ 2 \tilde{\lambda} + 2 \tilde{\sigma}} 
\end{equation}
where $\rho_{I 0} > 0 $ is a constant. One also obtains, for any
function $X(r(\tilde{\sigma}))$,
\[
X_{\tilde{\sigma}} = r X_r \; \; , \; \; \; 
X_{\tilde{\sigma} \tilde{\sigma}} = r^2 X_{r r} + r X_r
\]
\[
r^2 \left( X_{r r} + (\tilde{\Lambda}_r - \tilde{\lambda}_r)
\; X_r \right) \; = \; 
X_{\tilde{\sigma} \tilde{\sigma}} + 
(\tilde{\chi}_{\tilde{\sigma}} -
\tilde{\lambda}_{\tilde{\sigma}}) \; X_{\tilde{\sigma}}
\]
where the subscripts $\tilde{\sigma}$ denote $\tilde
{\sigma}-$derivatives and $\tilde {\chi} = \tilde {\Lambda} -
\tilde {\sigma} = \tilde {\lambda}^0 + (m - 1) \; \tilde
{\sigma} \;$. Equations (\ref{tld0rr}) -- (\ref{2tldLambdar22}),
written in terms of $\tilde{\sigma}$, now become
\begin{eqnarray}
\tilde{\lambda}^0_{\tilde{\sigma} \tilde{\sigma}} 
+ (\tilde{\chi}_{\tilde{\sigma}} 
- \tilde{\lambda}_{\tilde{\sigma}}) \; 
\tilde{\lambda}^0_{\tilde{\sigma}} & = & 
\sum_I \tilde{c}^{0 I} \; \rho_{I 0} \; e^{\phi^I 
+ 2 \tilde{\lambda} + 2 \tilde{\sigma}} \label{w0rr} \\
\tilde{\lambda}^i_{\tilde{\sigma} \tilde{\sigma}} 
+ (\tilde{\chi}_{\tilde{\sigma}} 
- \tilde{\lambda}_{\tilde{\sigma}}) \; 
\tilde{\lambda}^i_{\tilde{\sigma}} & = & 
\sum_I \tilde{c}^{i I} \; 
\rho_{I 0} \; e^{\phi^I + 2 \tilde{\lambda} 
+ 2 \tilde{\sigma}} \label{wirr} 
\end{eqnarray}
\begin{eqnarray}
\tilde{\lambda}^0_{\tilde{\sigma}} & = & \sum_I \frac {w^I_\pi}
{m} \; \rho_{I 0} \; e^{\phi^I + 2 \tilde{\lambda} 
+ 2 \tilde{\sigma}} + \frac{m - 1}{2} \; \left( 
e^{2 \tilde{\lambda}} - 1 \right) + \frac{r^2 {\cal B}}{2 m}
\label{wquad} \\
& & \nonumber \\
\tilde{\lambda}^0_{\tilde{\sigma}} 
- \tilde{\lambda}_{\tilde{\sigma}} & = & \sum_I \tilde{c}^I \;
\rho_{I 0} \; e^{\phi^I + 2 \tilde{\lambda} + 2 \tilde{\sigma}}
+ (m - 1) \; \left( e^{2 \tilde{\lambda}} - 1 \right)
\label{wsigma} \\
& & \nonumber \\
\Longrightarrow \; \; \; \; 
\tilde{\lambda}_{\tilde{\sigma}} & = & \sum_I \frac {\rho_{I 0}}
{m} \; \; e^{\phi^I + 2 \tilde{\lambda} + 2 \tilde{\sigma}} 
- \frac{m - 1}{2} \; \left( e^{2 \tilde{\lambda}} - 1 \right) 
+ \frac{r^2 {\cal B}}{2 m} \label{wq-sig}
\end{eqnarray}
where $r^2 \; {\cal B} = \sum_i \tilde
{\lambda}^i_{\tilde{\sigma}} \; \lambda^i_{\tilde{\sigma}} \;$.
The above equations thus describe stars whose constituents obey
the linear equations of state (\ref{eoswi}).


\vspace{4ex} 

\centerline{\bf 4. Asymptotic analysis : Singular solutions and
perturbations}

\vspace{2ex}

Consider the limit $r = e^{\tilde{\sigma}} \to \infty \;$.
Suitable ansatzes for $\tilde{\lambda}^0, \; \tilde
{\lambda}^i$, and $\tilde {\lambda}$ in this limit are given by
\begin{equation}\label{ansatz}
\tilde{\lambda}^0 = \tilde{s}^0 \; \tilde{\sigma} + \tilde{u}^0
\; \; , \; \; \;
\tilde{\lambda}^i = \tilde{s}^i \; \tilde{\sigma} + \tilde{u}^i
\; \; , \; \; \;
\tilde{\lambda} = \tilde{\lambda}_0 + \tilde{s} \;
\tilde{\sigma} + \tilde{u}
\end{equation}
where $\tilde{s}^0, \; \tilde{s}^i, \; \tilde{s}$, and $\tilde
{\lambda}_0$ are constants and $\tilde{u}^0, \; \tilde {u}^i$,
and $\tilde{u}$ are functions of $r \;$. Further constants
$\tilde {\lambda}^0_0$ and $\tilde {\lambda}^i_0$ could have
been added to $\tilde{\lambda}^0$ and $\tilde {\lambda}^i$ also
but, with no loss of generality, they have been set to zero. 
Writing
\begin{equation}\label{phiansatz}
\phi^I = q^I \; \tilde{\sigma} + y^I
\end{equation} 
it follows from equation (\ref{phii}) that $q^I$ and $y^I$ are
given by
\begin{eqnarray} 
w^I_\pi \; q^I & = & - \; (1 + w^I_\pi) \; \tilde{s}^0 
- 2 \; \eta^I w^I_\pi - \sum_i c^{i I} 
\; \tilde{s}^i \label{qi} \\
& & \nonumber \\
w^I_\pi \; y^I & = & - \; (1 + w^I_\pi) \; \tilde{u}^0 
- \sum_i c^{i I} \; \tilde{u}^i \label{yi} \; \; .
\end{eqnarray} 
Also, write $r^2 \; {\cal B} = \sum_i
\tilde{\lambda}^i_{\tilde{\sigma}} \; \lambda^i_{\tilde{\sigma}}
= {\cal B}_0 + 2 {\cal B}_1 + {\cal B}_2$ where
\[
{\cal B}_0 = \sum_i \tilde{s}^i \; s^i \; \; , \; \; \; 
{\cal B}_1 = \sum_i s^i \; \tilde{u}^i_{\tilde{\sigma}} 
\; \; , \; \; \;
{\cal B}_2 = \sum_i \tilde{u}^i_{\tilde{\sigma}} \; 
u^i_{\tilde{\sigma}} \;\; .
\]
In the limit $r \to \infty \;$, the $\tilde{\lambda}_0$ and the
$\tilde{\sigma}$ terms in the above equations give the leading
zeroth order asymptotic solutions. They give the singular
solutions of \cite{cpaper, pc07}. The functions $\tilde{u}$'s
are treated as perturbations and are used to obtain the first
order corrections to the leading asymptotic solutions. They give
the perturbations around the singular solutions.


\vspace{2ex} 

\centerline{\bf Zeroth order : Singular solutions}

\vspace{2ex}

Consider the equations of motion (\ref{w0rr}) -- (\ref{wq-sig})
and expand them to zeroth and first order in the functions
$\tilde {u}$'s. At zeroth order, equating the powers of $r$
gives 
\begin{equation}\label{qis}
2 + q^I + 2 \tilde{s} \; = \; \tilde{s} \; = \; 0 
\end{equation}
and, hence, $q^I = - 2 \;$. \footnote{ In general, one should
analyse equation (\ref{qi}) which determines $q^I$ and thus the
asymptotic behaviour of $\rho_I \;$. For a given set of values
for $w^I_\pi$ and $w^I \;$, some of the resulting $q^I$s may
lead to subdominant terms. Then the corresponding $\rho_I$s
become unimportant and effectively reduce ${\cal N} \;$. With no
loss of generality, we are assuming that $w^I_\pi$ and $w^I \;$
are such that $q^I = - 2$ for all $I$, thus all $\rho_I$s remain
important and ${\cal N}$ remains unreduced. \label{redN}} Now,
equation (\ref{qi}) becomes
\begin{equation}\label{q2i}
2 \; w^I_\pi \; (1 - \eta^I) = (1 + w^I_\pi) \; \tilde{s}^0 
+ \sum_i c^{i I} \; \tilde{s}^i \; \; . 
\end{equation}
Also, upto first order in the functions $\tilde{u}$'s, we have
\begin{eqnarray*}
r^2 (\rho_I e^{2 \lambda}) & = & \rho_{I 0} 
\; e^{2 \tilde{\lambda}_0 + y^I + 2 \tilde{u}} 
\; = \;  R_I \; (1 + y^I + 2 \tilde{u} + \; \cdots \; ) \\ 
& & \\
e^{2 \tilde{\lambda}} - 1 & = & e^{2 \tilde{\lambda}_0 
+ 2 \tilde{u}} - 1 \; = \; (e^{2 \tilde{\lambda}_0} - 1) 
+ e^{2 \tilde{\lambda}_0} \; (2 \tilde{u} + \; \cdots \; ) \\
& & \\
\tilde{\chi}_{\tilde{\sigma}} - \tilde{\lambda}_{\tilde{\sigma}}
& = & \alpha + (\tilde{u}^0_{\tilde{\sigma}} 
- \tilde{u}_{\tilde{\sigma}}) \; \; \; , \; \; \; \;
\alpha \; = \; m - 1  + \tilde{s}^0 - \tilde{s} 
\end{eqnarray*}
where $R_I = \rho_{I 0} \; e^{2 \tilde{\lambda}_0} \;$. At
zeroth order, equations (\ref{w0rr}) -- (\ref{wq-sig}) now give
\begin{eqnarray}
\alpha \; \tilde{s}^0 & = & \sum_I \tilde{c}^{0 I} \; R_I
\; \; \; , \; \; \; \; 
\alpha \; \tilde{s}^i \; = \; \sum_I \tilde{c}^{i I} \; R_I
\label{00i} \\
\tilde{s}^0 & = & \sum_I \frac {w^I_\pi}{m} \; R_I 
+ \frac{m - 1}{2} \; \left( e^{2 \tilde{\lambda}_0} - 1 \right)
+ \frac{{\cal B}_0}{2 m} \label{0quad} \\
\tilde{s}^0 - \tilde{s} & = & \sum_I 
\tilde{c}^I \; R_I + (m - 1) \; 
\left( e^{2 \tilde{\lambda}_0} - 1 \right) \label{0sigma} \\
& & \nonumber \\
\tilde{s} & = & \sum_I \frac{R_I}{m} 
- \frac {m - 1} {2} \; \left( e^{2 \tilde{\lambda}_0} 
- 1 \right) + \frac{{\cal B}_0}{2 m} \; \; , \label{quad} 
\end{eqnarray}
and equation (\ref{0sigma}) gives 
\begin{equation}\label{alpha}
\alpha \; = \; m - 1 + \tilde{s}^0 - \tilde{s} \; = \; 
\sum_I \tilde{c}^I \; R_I + (m - 1) \; 
e^{2 \tilde{\lambda}_0}  \; \; . 
\end{equation}
Equations (\ref{q2i}) -- (\ref{alpha}) constitute the equations
of motion at the leading zeroth order in the functions
$\tilde{u}$'s. They will give the singular solutions. 


\vspace{2ex} 

\centerline{\bf First order : perturbations around singular
solutions}

\vspace{2ex}

At first order in the $\tilde {u}$'s, equations (\ref{w0rr}) --
(\ref{wq-sig}) give
\begin{eqnarray*}
\tilde{u}^0_{\tilde{\sigma} \tilde{\sigma}} 
+ \alpha \tilde{u}^0_{\tilde{\sigma}} 
+ \tilde{s}^0 \; (\tilde{u}^0_{\tilde{\sigma}} 
- \tilde{u}_{\tilde{\sigma}}) & = &
\sum_I \tilde{c}^{0 I} \; R_I \; 
(y^I + 2 \tilde{u}) \label{10} \\
\tilde{u}^i_{\tilde{\sigma} \tilde{\sigma}} 
+ \alpha \tilde{u}^i_{\tilde{\sigma}} 
+ \tilde{s}^i \; (\tilde{u}^0_{\tilde{\sigma}} -
\tilde{u}_{\tilde{\sigma}}) & = &
\sum_I \tilde{c}^{i I} \; R_I \; (y^I + 2 \tilde{u}) \label{1i}
\end{eqnarray*}
\begin{eqnarray*}
\tilde{u}^0_{\tilde{\sigma}} & = &
\sum_I \frac{w^I_\pi}{m} \; R_I \; (y^I + 2 \tilde{u}) 
+ (m - 1) \; e^{2 \tilde{\lambda}_0} \; \tilde{u}
+ \frac{{\cal B}_1}{m} \label{1quad} \\
& & \nonumber \\
\tilde{u}^0_{\tilde{\sigma}} - \tilde{u}_{\tilde{\sigma}} 
& = & \sum_I \tilde{c}^I \; R_I \; 
(y^I + 2 \tilde{u}) + (m - 1) \; e^{2 \tilde{\lambda}_0} \; 
(2 \tilde{u}) \label{1sigma} \\
\tilde{u}_{\tilde{\sigma}} & = &
\sum_I \frac{R_I }{m} \; (y^I + 2 \tilde{u}) 
- (m - 1) \; e^{2 \tilde{\lambda}_0} \; \tilde{u}
+ \frac{{\cal B}_1}{m} \label{11quad} \; \; .
\end{eqnarray*}
Using the zeroth order results for the $\tilde{u}-$terms in the
right hand sides of the above equations, one obtains
\begin{eqnarray}
\tilde{u}^0_{\tilde{\sigma}} & = &
\sum_I \frac{w^I_\pi}{m} \; R_I \; y^I 
+ \left( 2 \tilde{s}^0 + m - 1 - \frac{{\cal B}_0} {m} \right)
\tilde{u} + \frac{{\cal B}_1}{m} \label{2quad} \\
& & \nonumber \\
\tilde{u}^0_{\tilde{\sigma}} - \tilde{u}_{\tilde{\sigma}} 
& = & \sum_I \tilde{c}^I \; R_I \; y^I 
+ 2 \alpha \tilde{u} \label{2sigma} \\     
\tilde{u}_{\tilde{\sigma}} & = &
\sum_I \frac{R_I }{m} \; y^I 
- \left(m - 1 + \frac{{\cal B}_0} {m} \right) \tilde{u}
+ \frac{{\cal B}_1}{m} \label{21quad} \; \; .
\end{eqnarray}
The $\tilde {u}^0_{\tilde {\sigma} \tilde {\sigma}}$ and the
$\tilde {u}^i_{\tilde {\sigma} \tilde {\sigma}}$ equations now
become
\begin{eqnarray}
\tilde{u}^0_{\tilde{\sigma} \tilde{\sigma}} + \alpha
\tilde{u}^0_{\tilde{\sigma}} & = & \sum_I (\tilde{c}^{0 I} 
- \tilde{s}^0 \tilde{c}^I) \; R_I \; y^I \label{20} \\
\tilde{u}^i_{\tilde{\sigma} \tilde{\sigma}} + \alpha
\tilde{u}^i_{\tilde{\sigma}} & = & \sum_I (\tilde{c}^{i I} 
- \tilde{s}^i \tilde{c}^I) \; R_I \; y^I \label{2i}
\end{eqnarray}
Equations (\ref{yi}) and (\ref{2quad}) -- (\ref{2i}) constitute
the equations of motion at first order in the functions
$\tilde{u}$'s. Their solutions give the perturbations around the
singular solutions.


\vspace{2ex} 

\centerline{\bf Significance of the perturbations}

\vspace{2ex}

We now outline briefly the significance of the singular
solutions and the perturbations around them. In section {\bf 2},
we have briefly mentioned a few of these aspects. See \cite
{pc07} for complete details.

The singular solutions and the perturbations around them can be
used, among other things, as indicators of the stability of
stars. A star, with its constituents obeying linear equations of
state, is enclosed in a box of radius $r_*$ so as to prevent the
evaporation of its constituents and to make its total mass
finite. The mass of the star is then given, upto constant
numerical factors, by the mass function $M(r_*)$ evaluated at
$r_* \;$. The singular solutions give the scaling relations
between mass, radius, and other quantities of the stars. The
singular solutions are of no relevance when the stars are
unstable but, when stable, these solutions give the scaling
relations in the limit of large central density or large radius
of the stable stars.

The behaviour, namely damped oscillatory or monotonic non
oscillatory, of the perturbations around the singular solutions
leads to the corresponding behaviour for the mass -- central
density profile in the asymptotic regions. Considering a series
of equilibria, Chavanis obtains the mass -- central density
profile in the asymptotic and non asymptotic regions. He shows
that equilibrium configurations become unstable beyond the first
maximum in this profile. This is also shown to correspond to
conditions for nonlinear dynamic stability. In those higher
dimensional cases where the behaviour of the perturbations is
monotonic non oscillatory, Chavanis obtains the mass -- central
density profile in the non asymptotic regions also and shows
that the entire profile is monotonically increasing and
asymptotes to the constant value line given by the singular
solutions.

In this paper, assuming linear equations of state, we will
obtain singular solutions and perturbations around them.
Enclosing the star in a box of radius $r_*$, the mass -- central
density profile and its behaviour in the asymptotic regions can
also be obtained from the perturbations. However, in the case of
horizonless objects of interest here, nothing is known
rigorously about the nature and properties of their
constituents. Hence, we are unable to carry out the analogs of
the analyses mentioned in the previous paragraph. Namely, since
the detailed properties of the constituents are not known, we
are unable to obtain the profile in the non asymptotic regions,
and to study the stability properties around the maxima.

We note below a few useful points.

\vspace{2ex} 

{\bf (1)} 
The expressions $\frac{M(r)} {r^{m - 1}} = 1 - e^{- 2 \tilde
{\lambda}}$ and $\tilde {\lambda} = \tilde {\lambda}_0 + \tilde
{u}$ imply that, in the limit of large $r$,
\begin{equation}\label{mru}
\frac{M(r)} {r^{m - 1}} \; = \; 1 - e^{- 2 \tilde{\lambda}_0} 
+ 2 \; e^{- 2 \tilde{\lambda}_0} \; \tilde{u} + \cdots \; .
\end{equation}
The first two terms correspond to the singular solutions and the
$\tilde{u}$ term describes the perturbations in the mass
function. Similarly, $y^I$ describes the perturbations in the
density $\rho^I \;$.

\vspace{2ex}

{\bf (2)} 
The radius $r_*$ of a spherically symmetric star, defined to be
given by $\Pi (r_*) = 0$, is infinite when it is made up of
perfect fluids with linear equations of state, see \cite{rs} for
a derivation. Hence, let the star be enclosed in a box of radius
$r_* \;$ which will render its mass 
\[
M(r_*) \simeq \int^{r_*} \; dr \; M_r
\]
finite where $M_r$ is given in equation (\ref{mr}). Then,
following the analysis of \cite{pc07}, the perturbations around
the singular solutions can be used to obtain the mass -- central
density profile of the star in the asymptotic limit of large
central density or large radius.

Let $x_{ch} \propto \sqrt{\rho_c} \; r_*$ be a measure of
central density $\rho_c$, and let $y_{ch} = \frac {M(r_*)}
{r_*^{m - 1}}$ be a measure of the mass of the star. By detailed
analysis of the equations, and incorporating the properties of
the constituents of the star, Chavanis obtains the mass --
central density profile in both the asymptotic and non
asymptotic regions and finds that : {\bf (i)} As $x_{ch}$
increases from zero to $\infty$, $\; y_{ch}$ increases from zero
to a (first) maximum $y_1$ at $x_1$, thereafter exhibits damped
oscillations, asymptoting to a value $y_s \;$. {\bf (ii)} The
behaviour for large values of $x_{ch}$ can be seen from the
singular solutions and the asymptotic perturbations around
them. {\bf (iii)} The solutions are unstable beyond the first
maximum which is at $(x_{ch}, \; y_{ch}) = (x_1, \; y_1) \;$.

As a consequence, one has the following. For a given value of
central density, the radius $r_*$ must be $< r_{1 *}$ where
$r_{1 *}$ corresponds to $x_1$. The mass of the star must then
be less than $(y_1 \; r_{1 *}^{m - 1}) \;$. A more massive star
will be unstable and will collapse.

\vspace{2ex}

{\bf (3)} 
For $D = m + 2$ dimensional stars, $n_c = 0$ and ${\cal N} = 1$
in our notation, Chavanis finds that if $m \ge m_{cr} \sim 9$
then $y_{ch}$ increases monotonically from zero to $y_s$,
effectively making $x_1$ infinite and $y_1 = y_s \;$. See Figure
23, and also Figures 20 -- 22, in \cite{pc07}. The asymptotic
perturbations around the corresponding singular solutions
exhibit monotonic behaviour with no oscillations.

As a consequence, one has the following. For $m \ge 9$, $\; x_1$
is effectively infinite. Then, for a given value of central
density, $r_{1 *}$ is infinite which makes the upper limit $(y_s
\; r_{1 *}^{m - 1})$ on the mass of the star also infinite.
Hence, a star can be arbitrarily massive and stable when $m \ge
9 \;$. Also, in the limit of large central density or large
radius, one obtains the scaling relation $M(r_*) \sim r_*^{m -
1}$ which shows that the sizes of the large stable stars scale
as their Schwarzschild radii.

\vspace{2ex}

{\bf (4)} 
In \cite{k13}, we generalised this study to $D = n_c + m + 2$
dimensional stars, with $n_c$ toroidal directions, made up of
${\cal N} > 1$ number of perfect fluids where pressures are
isotropic along the non compact spatial directions. Stars in M
theory correspond to specific values of $n_c$ and ${\cal N}
\;$. We found that, even in these generalised cases, $m \ge 9$
is required for stability.


\vspace{4ex} 

\centerline{\bf 5. Singular solutions and asymptotic
perturbations}

\vspace{2ex}

Taking the pressures inside the stars to be anisotropic along
the radial and the spherical directions of the non compact
space, and hence taking the anisotropy parameters $\eta^I \ne
0$, we now obtain singular solutions to the equations of motion
and asymptotic perturbations around them, generalising those
given in \cite{cpaper, pc07, k13}. These solutions may be
obtained for any general set of values for $n_c$, $\; m$, $\;
{\cal N}$, $\; w^I_\alpha$ and $\; w^I_\pi$. However, such a
generality is neither illuminating nor needed for our purposes
here. Hence we will present only three cases which are
illustrative and are also of direct interest.


In this section we will consider $(m + 2)$ dimensional stars
made up of a single fluid. In the next section we will consider
two examples of stars in M theory : In one, the stars are made
up of a stack of $M 2$ or $M 5$ branes. In another, the stars
are made up of four stacks of intersecting branes, two stacks
each of $M 2$ and $M 5$ branes, see footnote {\bf \ref{ftmstar}
}.

\vspace{4ex}

\centerline{\bf ${ \mathbf (m + 2) }$ dimensional stars with 
${ \mathbf {\cal N} = 1 }$}

\vspace{2ex}

Consider $(m + 2)$ dimensional stars made up of a single fluid.
Then $n_c = 0$ and ${\cal N} = 1 \;$. The tildes and the
$I-$scripts on various quantities are now unnecessary. Hence we
omit them.

The fluid is assumed to have anisotropic pressures in general.
Its radial and the transverse spherical pressures, and the
anisotropy parameter $\eta \;$, are given by
\[
\Pi = w_\pi \; \rho \; \; , \; \; \; 
p = w \; \rho \; \; , \; \; \; 
\eta = \frac {m} {2} \; \left( \frac {w_\pi - w} {w_\pi} \right)
\; \; .
\]
Also, $c^0 = 1 + w + c$ and $c^a = c = \frac {w_\pi - 1} {m}$,
and hence
\[
m \; c^0 = (m - 1) \; (1 + w_\pi) + 2 \; w_\pi \; (1 - \eta) 
\; \; .
\]
We now write down the leading order asymptotic solutions to the
equations of motion and perturbations around them. We have $s =
0$ and $q = - 2 \;$. The zeroth order equations (\ref{q2i}) --
(\ref{alpha}) then give the following relations.
\begin{eqnarray*}
s^0 & = & \frac {2 \; w_\pi \; (1 - \eta) } {1 + w_\pi}
\; \; \Longrightarrow \; \; \; 
\alpha \; = \; m - 1 + s^0 \; = \; 
\frac {m \; c^0} {1 + w_\pi} \\
& & \\
R & = & \frac {\alpha \; s^0} {c^0} \; = \; 
\frac {2 \; m \; w_\pi \; (1 - \eta) } {(1 + w_\pi)^2} \\
& & \\
(m - 1) \; e^{2 \lambda_0} & = & \alpha - c \; R \; = \; 
\frac {{\cal D}} {(1 + w_\pi)^2} \\
& & \\
{\cal D} & = & (m - 1) \; (1 + w_\pi)^2 
+ 4 \; w_\pi \; (1 - \eta) \; \; .
\end{eqnarray*}
The above expressions describe the singular solutions. For
example, they give 
\[
e^{2 \lambda^0} \; \simeq \; r^{2 s^0} \; \; , \; \; \; 
e^{2 \lambda} \; \simeq \; 1 + \frac {4 w_\pi (1 - \eta)} 
{(1 + w_\pi)^2} \; \; , \; \; \; 
\]
\[
\rho \; \simeq \; \frac {\rho_0} {r^2} 
= \frac {2 m (m - 1) w_\pi (1 - \eta)} {r^2 {\cal D}}
\; \; . 
\]
Note that, from $R \propto \rho_0 > 0$ and $w_\pi > 0 \;$, it
follows that $\eta < 1 \;$ and, hence, that $m c^0$, $\; s^0$,
$\; \alpha$, and ${\cal D}$ are all positive. The mass function
can be obtained from equation (\ref{mru}) and, using the above
expression for $e^{2 \lambda_0} \;$, it is given by
\begin{equation}\label{mrum+2}
\frac{M(r)} {r^{m - 1}} \; = \; \frac {2} {{\cal D}} \; 
\left(2 \; w_\pi \; (1 - \eta) + (m - 1) \; (1 + w_\pi)^2
\; u + \cdots \right) \; \; .
\end{equation}
The mass of the star $M(r_*)$ enclosed in a box of large radius
$r_*$ is then given by
\[
M(r_*) \; \simeq \; \frac {4 w_\pi (1 - \eta)} {{\cal D}}
\; r_*^{m - 1} \; \; . 
\]

\vspace{2ex} 

Consider now the first order equations of motion (\ref{yi}) and
(\ref{2quad}) -- (\ref{20}). Their solutions will give the
asymptotic perturbations around the singular solutions. Equation
(\ref{yi}) gives, upon using the expressions for $s^0$ and $R
\;$,
\[ 
s^0 \; y \; = \; - \; 2 \; (1 - \eta) \; u^0 \; \; , \; \; \;
R \; y \; = \; - \; 2 \; (1 - \eta) \; \alpha \; \frac {u^0}
{c^0} \; \; .
\]
Equations (\ref{2quad}) and (\ref{21quad}) give
\begin{eqnarray*}
u^0_\sigma & = & - s^0 \; u^0 + (m - 1 + 2 s^0) \; u \\
& & \\
u_\sigma & = & - \frac{s^0}{w_\pi} \; u^0 - (m - 1) \; u
\end{eqnarray*}
from which it follows, after a little algebra, that both $u^0$
and $u$ obey the same equation given by
\begin{equation}\label{*1}
(*)_{\sigma \sigma} + \alpha \; (*)_\sigma + \frac {2 \; 
(1 - \eta) \; {\cal D}} {(1 + w_\pi)^2} \; (*) \; = \; 0 
\end{equation}
where $(*) = u^0$ or $u \;$. It is straightforward to show that
equation (\ref{20}) also gives the above equation for $u^0 \;$.

The solutions to equation (\ref{*1}) are of the form $(*) \sim
e^{k \sigma}$ where
\[
k \; = \; \frac {- \alpha \pm \sqrt{\Delta}} {2} 
\; \; , \; \; \; \;
\Delta = \alpha^2 - \frac {8 \; (1 - \eta) \; {\cal D}} 
{(1 + w_\pi)^2} \; \; .
\]
If $\Delta < 0$ then, in an obvious notation, $k = - \; k_{re}
\pm i \; k_{im} \;$ with $k_{re} > 0 \;$. The solutions for
$(*)$ are then oscillatory and, since $\sigma = ln \; r \;$,
they may be written as
\[
(*) = \frac {A_*} {r^{k_{re}}} \; \; 
Sin \; (k_{im} \; ln \; r + B_*)
\]
where $A_*$ and $B_*$ are integration constants. If $\Delta > 0$
then $\alpha > \sqrt{\Delta}$ since $1 - \eta > 0$ and ${\cal D}
> 0 \;$ and, in an obvious notation, $k = - k_1 \pm k_2$ with
$k_1 > k_2 > 0 \;$. Hence the solutions for $(*)$ are non
oscillatory and they may be written as
\[
(*) = \frac {A_*}  {r^{k_1 - k_2}} \; 
\left( 1 + \frac {B_*}  {r^{2 k_2}} \right) \; \; .
\]
If $\Delta = 0$ then $k_2 = 0 \;$ and the solutions are $(*) =
\frac {A_*} {r^{k_1} } \; (ln \; r + B_*)$ and are non
oscillatory. Considering the star enclosed in a box of radius
$r_*$, introducing Milne variables, and following the analysis
of Chavanis given in \cite {pc07}, one can now obtain the mass
-- central density profile in the asymptotic limit of large
central density or large radius. The asymptotic behaviour of
this profile is similar to that of the perturbations: it is
oscillatory if $\Delta < 0$, and is non oscillatory if $\Delta
\ge 0 \;$.


Comparing with the work of Chavanis in \cite{pc07}, we note that
the solutions for $(*)$ given above when $\Delta < 0$ and when
$\Delta \ge 0$ will lead to the analogs of equations (174) and
(175) in \cite{pc07}. The resulting mass -- central density
profile for $M(r_*)$ will lead to the analogs of the asymptotic
parts of Figure 23 in \cite{pc07}. The numerical and the
analytical studies leading to the analogs of the non asymptotic
initial parts of that Figure are beyond the scope of the present
paper and hence, although important, are not attempted here. If
one assumes that these non asymptotic initial parts remain
qualitatively the same for the anisotropic case also, then one
may conclude that a star can be arbitrarily massive and stable
when $\Delta \ge 0 \;$; and that the singular solutions describe
its mass -- radius relations in the limit of large central
density or large radius.

\vspace{2ex}

We now study the conditions under which $\Delta < 0$ and $\Delta
\ge 0 \;$. Let $\Delta = \frac {\hat{\Delta}} {(1 + w_\pi)^2}
\;$. It then follows that
\begin{equation}\label{kq0}
k \; = \; \frac {- m \; c^0 \pm \sqrt{\hat{\Delta}}} 
{2 \; (1 + w_\pi)} \; \; , \; \; \; \; 
\hat{\Delta} \; = \; A \; w_\pi^2 + 2 B \; w_\pi + C
\end{equation}
where, after some algebra, one obtains  
\begin{eqnarray*}
A & = & (m - 3 + 2 \eta)^2 \\ 
& & \\
B & = & (m + 1 - 2 \eta) \; (m - 9 + 8 \eta) \\
& & \\
C & = & (m - 1) \; (m - 9 + 8 \eta) \; \; , \\ 
& & \\
A C - B^2 & = & 32 \; (1 - \eta)^2 \; (m - \eta) \; 
(m - 9 + 8 \eta) \; \; .
\end{eqnarray*}
Consider now the sign of $\Delta \;$. It is same as that of
$\hat {\Delta} \;$. Using the quadratic expression for $\hat
{\Delta}$ given above, it can be seen that if $A C - B^2 < 0$
then $\hat{\Delta}$ can be negative for a range of values for
$m$, $\; \eta$, and $w_\pi \;$. For example, if $m = 2$ and
$\eta = 0$ then $A C - B^2 < 0 \;$. Then $\hat{\Delta} = w_\pi^2
- 42 w_\pi - 7$ and is negative, for example, for $0 < w_\pi < 1
\;$. If $A C - B^2 \ge 0 \;$ then $\hat{\Delta} \ge 0$ always.

\vspace{2ex} 

In the isotropic case, $\eta = 0$ and the above expressions
reduce to those given in \cite{pc07}. Then $\hat{\Delta} \ge 0$
if $m \ge 9$ and, depending on the value of $w_\pi = w$, $ \;
\hat{\Delta}(w) \ge 0$, if $m \ge m_{cr}(w) \sim 9 \;$. In the
anisotropic case, $\eta$ is non vanishing. Noting that $\eta <
1$ and $m \ge 2 \;$, it follows from the above expressions that
$A C - B^2 \ge 0 \;$ if the values of $\eta$ and $\frac {w}
{w_\pi}$ lie in the range given by
\begin{equation}\label{etam+2}
\frac{9 - m} {8} \le \eta < 1
\; \; \; \longleftrightarrow \; \; \; 
\frac {5 m - 9} {4 \; m} \ge \frac {w} {w_\pi} > \frac {m - 2}
{m} \; \; ;
\end{equation}
then $\hat {\Delta} \ge 0$, hence $\Delta \ge 0$, and the
solutions are non oscillatory. The dependence of these ranges on
$w_\pi$ can also be incorporated, but is superfluous for our
present purposes of showing that anisotropy can lead to
asymptotic non oscillatory solutions. Note that the isotropic
case $\eta = 0$, equivalently $w = w_\pi \;$, is included in the
ranges given above only when $m \ge 9 \;$. For lower values of
$m$, (for example, for $m = 2$ which corresponds to four
dimensional spacetime) certain amount of anisotropy ( $\eta \ge
\frac {7} {8} \;$) is needed to obtain the non oscillatory
behaviour of the asymptotic perturbations around the singular
solutions.


\vspace{4ex}

\centerline{\bf 6. Stars in M theory} 

\vspace{2ex}

In this section we will analyse stars in M theory. In section
{\bf 2}, we have described briefly M theory stars and how the
necessary equations of state are obtained. We now proceed with
the analysis.

The spacetime is eleven dimensional in M theory, having $n_c$
dimensional compact toroidal space and $(m + 2)$ dimensional non
compact spacetime where $n_c + m = 9 \;$. The M theory stars are
taken to be made up of ${\cal N}$ stacks of $M 2$ and $M 5$
branes, intersecting according to the BPS rules. We take the
spatial directions of the brane worldvolumes to be toroidal and
assume necessary isometries. These intersecting branes can be
modelled by ${\cal N}$ number of seperately conserved energy
momentum tensors \cite{alg, aeh, aeh2, aeh3}. Hence, the present
formalism can be applied to the corresponding stars.

M theory has U duality symmetries. As shown in \cite{k07}, see
also footnote {\bf \ref{ftk07U}}, they lead to a relation among
the components $(p_{\alpha I}, \; \Pi_I)$ of the energy momentum
tensor $T_{M N (I)}$ which is given by
\[ 
p_{\parallel I} \; = \; \Pi_I + p_{0 I} + p_{\perp I} 
+ m \; (p_I - p_{\perp I}) 
\]
where $p_{\parallel I}$ and $p_{\perp I}$ are the pressures
along the directions that are parallel and transverse to the
worldvolume of the $I^{th}$ stack of branes. The above relation
is a consequence of U duality symmetries and, therefore, must
always be valid independent of the details of the equations of
state. Also, since the sphere directions are transverse to the
branes, it is natural to set $p_I = p_{\perp I} \;$. 

Note that, for stars in M theory, there is no compelling reason
to take the pressures $\Pi_I$ and $p_I$ to be equal. Indeed, in
the case of charged intersecting black branes in M theory, these
pressures are not equal although the relation $p_I = p_{\perp I}
\;$ and the U duality relation above are obeyed. Hence we assume
that, in general, $\Pi_I \ne p_I$ for stars in M theory.
Setting $p_{\perp I} = p_I \;$ and $p_{0 I} = - \rho_I \;$, the
U dulaity relation now becomes
\begin{equation}\label{ups} 
p_{\parallel I} \; = \; \Pi_I - \rho_I + p_I \; \; . 
\end{equation} 
Consider the linear equations of state given by (\ref{eoswi}).
Let $p_{\parallel I} = w^I_\parallel \; \rho_I \;$ and $p_{\perp
I} = w^I_\perp \; \rho_I \;$ where $w^I_\perp = w^I$ since
$p_{\perp I} = p_I \;$. The U duality relation then gives
\[
w^I_\parallel = w^I_\pi - 1 + w^I \; \; . 
\]

Now consider the coefficients $\tilde{c}^{i I} = - w^I_i + w^I +
\tilde{c}^I$ defined in equation (\ref{tldcalpha}). Note that
$w^I_i = w^I_\parallel$ if $\; i \in \; \parallel_I \;$, namely
if $x^i$ is a worldvolume coordinate of the $I^{th}$ stack of
branes; otherwise, $w^I_i = w^I_\perp \;$. It then follows from
$w^I_\perp = w^I$ and the U duality relation for $w^I_\parallel$
given above, that the corresponding coefficients $\tilde
{c}^{\parallel I}$ and $\tilde {c}^{\perp I}$ are given by
\begin{equation}\label{uc}
\tilde{c}^{\parallel I} = - w^I_\pi + 1 + \tilde{c}^I 
=  (1 - m) \; \tilde{c}^I  
\; \; , \; \; \;
\tilde{c}^{\perp I} = \tilde{c}^I = \frac{w^I_\pi - 1}{m}
\; \; . 
\end{equation}
Hence, for any $a^i$ with $\tilde{a}^i = a^i + \frac {a^c} {m}
\;$ and $a^c = \sum_i a^i \;$, it follows that
\begin{equation}\label{cai}
\sum_i c^{i I} \; \tilde{a}^i \; = \; 
\sum_i \tilde{c}^{i I} \; a^i \; = \; \tilde{c}^I \; a^c 
\; - \; m \; \tilde{c}^I \; \sum_{i \in \parallel_I} a^i \; \; .
\end{equation}
In passing, we note that the relation $\tilde{c}^{\parallel I} =
(1 - m) \; \tilde{c}^{\perp I} \;$ is same as that obtained in
the corresponding isotropic cases studied in \cite{k13}.
Consequently, the resulting equations for M theory stars will
have very similar structure in both the isotropic and the
anisotropic cases. With $w^I_\alpha$ and $\tilde{c}^{\alpha I}$
specified for intersecting branes, we now consider two examples
of stars in M theory.


\vspace{4ex} 

\centerline{\bf M theory stars made up of ${ \mathbf M 2 }$ or
${ \mathbf M 5 }$ branes}

\vspace{2ex}

Consider stars in M theory made up of a stack of $M 2$ or $M 5$
branes. Then ${\cal N} = 1$, $\; n_c = 2$ or $5$, and $m = 7$ or
$4 \;$. The $I-$scripts on various quantities are now
unnecessary and, hence, we omit them. Also, we will first write
the solutions in a form applicable for ${\cal N} = 1$ and for
any values of $n_c$, $\; m$, and $\tilde {c}^i \;$; and then, at
the end, specialise to the case of $M 2$ or $M 5$ brane stars.

We now write down the leading order asymptotic solutions to the
equations of motion and perturbations around them. We have
$\tilde{s} = 0$ and $q = - 2 \;$. The zeroth order equations
(\ref{q2i}) -- (\ref{0sigma}) then give the following relations.
They describe the analogs of singular solutions in this context.
\begin{eqnarray*} 
2 \; w_\pi \; (1 - \eta) & = &  (1 + w_\pi) \; \tilde{s}^0 
+ \sum_i c^i \; \tilde{s}^i \label{25q2i} \\
& & \nonumber \\
\alpha \; \tilde{s}^0 & = & \tilde{c}^0 \; R
\; \; \; , \; \; \; \; 
\alpha \; \tilde{s}^i \; = \; \tilde{c}^i \; R
\label{2500i} \\
& & \nonumber \\
\tilde{s}^0 & = & \frac {w_\pi}{m} \; R
+ \frac{m - 1}{2} \; \left( e^{2 \tilde{\lambda}_0} - 1 \right)
+ \frac{{\cal B}_0}{2 m} \label{250quad} \\
& &  \nonumber \\
\tilde{s}^0 & = & \tilde{c} \; R + (m - 1) \; 
\left( e^{2 \tilde{\lambda}_0} - 1 \right) \; \; .  \label{250sigma}
\end{eqnarray*}
Using these relations, one obtains
\begin{equation}\label{rsis0}
R = \alpha \; \frac {\tilde{s}^0} {\tilde{c}^0} 
\; \; , \; \; \; 
\tilde{s}^i = \tilde {c}^i \; \frac {\tilde{s}^0} {\tilde{c}^0}
\; \; , \; \; \; 
\tilde{s}^0 = \frac {2 \; w_\pi \; (1 - \eta)} {(1 + w_\pi) \;
(1 + \gamma)} \\
\end{equation}
where $\gamma = \frac {\sum_i \tilde{c}^i \; c^i} { (1 + w_\pi)
\; \tilde{c}^0} \;$; and, after some algebra,
\begin{equation}\label{l0}
(m - 1) \; e^{2 \tilde{\lambda}_0} \; = \; \alpha - \tilde{c} \;
R \; = \; \frac{\alpha}{1 + \gamma} \; \left( \frac {{\cal D}}
{m \; \tilde{c}^0 \; (1 + w_\pi)} + \gamma \right) \\
\end{equation}
where ${\cal D} = (m - 1) \; (1 + w_\pi)^2 + 4 \; w_\pi \; (1 -
\eta) \;$. Note that, from $R \propto \rho_0 > 0$ and $w_\pi > 0
\;$, it follows that $\eta < 1 \;$; and, hence, that $m \tilde
{c}^0$, $\; \gamma$, $\; \tilde{s}^0$, $\; \alpha$, and ${\cal
D}$ are all positive. Note also that the effects of the compact
toroidal space appear through the parameter $\gamma \;$ alone.

Consider the first order equations of motion (\ref{yi}) and
(\ref{2quad}) -- (\ref{2i}). Their solutions will give the
asymptotic perturbations around the singular solutions. Upon
using $\tilde{s}^i = \tilde {c}^i \; \frac {\tilde{s}^0}
{\tilde{c}^0} \;$, equations (\ref{20}) and (\ref{2i}) give
\[
F^i_{\tilde{\sigma} \tilde{\sigma}} 
+ \alpha \; F^i_{\tilde{\sigma}} \; = \; 0 
\] 
where $F^i = \tilde{u}^i - \tilde{c}^i \; \frac {\tilde{u}^0}
{\tilde{c}^0} \;$. Although it follows that $F^i = F^i_1 \; e^{-
\alpha \tilde{\sigma}} \;$ in general, we will set the
integration constants $F^i_1$ to zero. This gives $\tilde{u}^i =
\tilde {c}^i \; \frac {\tilde{u}^0} {\tilde{c}^0} \;$, which we
now use in the remaining equations (\ref{yi}) and (\ref{2quad})
-- (\ref{20}).

Equation (\ref{yi}) gives, upon using equations (\ref{rsis0})
for $\tilde{s}^0$ and $R \;$,
\[ 
\tilde {s}^0 \; y \; = \; - \; 2 \; (1 - \eta) \; \tilde {u}^0
\; \; , \; \; \; 
R \; y \; = \; - \; 2 \; (1 - \eta) \; \alpha \; 
\frac {\tilde {u}^0} {\tilde {c}^0} \; \; . 
\]
Equation (\ref{2quad}) gives, after some manipulations,
\footnote {We used the expressions ${\cal B}_0 = \gamma (1 +
w_\pi) \frac {(\tilde {s}^0)^2} {\tilde{c}^0}$, $\; {\cal B}_1 =
\gamma (1 + w_\pi) \left( \frac {\tilde {s}^0} {\tilde{c}^0}
\right) \tilde {u}^0_{\tilde {\sigma}} \;$, and $\frac {m \;
c^0} {1 + w_\pi} = \alpha + \gamma s^0 \;$, which can all be
derived easily.}
\[
\tilde{u}^0_{\tilde{\sigma}} \; = \; 
- \; (1 + \gamma) \; \tilde{s}^0 \; \tilde{u}^0
+ (m - 1 + (2 + \gamma) \; \tilde{s}^0 
) \; \tilde{u} \; \; .
\]
Equation (\ref{2sigma}) gives straightforwardly
\[
\tilde{u}^0_{\tilde{\sigma}} - \tilde{u}_{\tilde{\sigma}} 
\; = \; - \; 2 \; (1 - \eta) \; \alpha \; \tilde{c} \; \frac
{\tilde {u}^0} {\tilde {c}^0} + 2 \; \alpha \; \tilde{u} \; \; .
\]
After a little algebra, it follows from these two equations that
both $u^0$ and $u$ obey the same equation which we write as
\begin{equation}\label{*m25}
(*)_{\tilde{\sigma} \tilde{\sigma}} + \alpha \;
(*)_{\tilde{\sigma}}
+ 2 \; (1 - \eta) \; (\alpha - \tilde{c} \; R) \; (*) \; = \; 0
\end{equation}
where $(*) = u^0$ or $u \;$ and $(\alpha - \tilde{c} \; R) \;$
is given in equation (\ref{l0}). Equation (\ref{20}) gives the
equation for $u^0$ straightforwardly in the above form.

The solutions to equation (\ref{*m25}) are of the form $(*) \sim
e^{k \sigma}$ where
\[
k \; = \; \frac {- \alpha \pm \sqrt{\Delta}} {2} 
\; \; , \; \; \; \;
\Delta = \alpha^2 - 8 \; (1 - \eta) \; (\alpha - \tilde{c} \; R)
\; \; .
\]
As explained in the case of $(m + 2)$ dimensional star, it is
important to study the sign of $\Delta \;$ since it determines
whether the solutions for $(*)$ are oscillatory or not. This
behaviour of the perturbations $(*)$ will lead to corresponding
asymptotic behaviour in the mass -- central density profile of a
star enclosed in a box. Writing
\[
\Delta \; = \; \frac {\alpha \; (\hat{\Delta} + \hat{\delta})}
{m \; \tilde{c}^0 \; (1 + w_\pi) \; (1 + \gamma)} \; \; , 
\]
it can be shown after a long but straightforward algebra that
$\hat{\Delta}$ and $\hat{\delta}$ are given by equation
(\ref{kq0}) and
\begin{equation}\label{q1}
\hat{\delta} \; = \; m \; (m - 9 + 8 \eta) \; 
\sum_i \tilde{c}^i \; c^i \; \; . 
\end{equation}

Consider the sign of $\Delta \;$. It is same as that of $(\hat
{\Delta} + \hat {\delta}) \;$. If $\tilde{c}^i = 0$ for all $i$
then $\hat {\delta} = \gamma = 0 \;$ and the earlier analysis
given below equation (\ref{kq0}) applies directly. Now let
$\tilde{c}^i$ do not all vanish. Then $\sum_i \tilde{c}^i c^i >
0 \;$, see footnote {\bf \ref{untilde}}. It then follows that
$\hat{\delta} \ge 0$ if $m - 9 + 8 \eta \ge 0 \;$. Hence, noting
that $\eta < 1$, one has $\hat{\delta} \ge 0$ if the
inequalities (\ref{etam+2}) are satisfied. Interestingly, this
is precisely the condition that also ensures that $\hat{\Delta}
\ge 0 \;$ but we do not know a simple reason, if any, for this
coincidence. We thus have that $\Delta \ge 0 \;$ and the
solutions are non oscillatory if the values of the anisotropy
parameter $\eta$ and, equivalently, of $\frac {w} {w_\pi}$ lie
in the ranges given in equation (\ref{etam+2}). Also, note that
the value $\eta = 0$, equivalently $w = w_\pi \;$, is included
in these ranges only when $m \ge 9 \;$. Otherwise certain amount
of anisotropy is needed along the non compact spatial dimensions
to obtain the non oscillatory behaviour of the perturbations
around the leading order asymptotic solutions.

The solutions obtained above are for ${\cal N} = 1$ and are
applicable for any values of $n_c$, $\; m$, and $\tilde{c}^i
\;$. In particular, they are also applicable to stars in M
theory made up of a stack of $M 2$ or $M 5$ branes for which
$n_c = 2$ or $5$, $\; m = 7$ or $4 \;$, and $\tilde {c}^i =
\tilde {c}^\parallel = (1 - m) \; \tilde{c}$ as given in
equation (\ref{uc}). It then follows easily that
\[
\sum_i \tilde{c}^i c^i \; = \; \frac {n_c \; m} {n_c + m} \;
(\tilde {c}^\parallel)^2 \; = \; A_{n_c} \; (w_\pi - 1)^2
\]
where $A_{n_c} = \frac{8}{7}$ for $M 2$ branes and $=
\frac{5}{4}$ for $M 5$ branes.


\vspace{4ex} 

\centerline{\bf M theory stars made up of ${ \mathbf {\cal N} =
4 } \;$, $\; { \mathbf 2 2' 5 5' }$ intersecting branes} 

\vspace{2ex}

Consider stars in M theory made up of two stacks each of $M 2$
and $M 5$ branes intersecting according to the BPS rules whereby
two stacks of five branes intersect along three common spatial
directions; a stack each of two branes and five branes intersect
along one common spatial direction; and two stacks of two branes
intersect along zero common spatial direction. Let $2 2' 5 5' :
(12, 34, 13567, 24567)$ denote the configuration of the
intersecting branes and indicate the spatial worldvolume
directions of the four stacks of branes. Now ${\cal N} = 4$, $\;
n_c = 7$, and $m = 2 \;$. Also, the coefficients $\tilde{c}^{i
I}$ are given by equations (\ref{uc}). Thus,
\[
\tilde{c}^{\parallel I} = - \; \tilde{c}^{\perp I} 
= - \; \tilde{c}^I 
= - \; \frac {w^I_\pi - 1} {2} \; \; .
\]
Note that equation (\ref{cai}) gives, with $S^c = \sum_i s^i$
and $U^c = \sum_i u^i \;$, 
\begin{eqnarray}
\sum_i c^{i I} \; \tilde{s}^i & = &
\sum_i \tilde{c}^{i I} \; s^i \; = \; \tilde{c}^I \; S^c 
\; - \; m \; \tilde{c}^I \; \sum_{i \in \parallel_I} s^i 
\label{cisi} \\
& & \nonumber \\
\sum_i c^{i I} \; \tilde{u}^i & = &
\sum_i \tilde{c}^{i I} \; u^i \; = \; \tilde{c}^I \; U^c 
\; - \; m \; \tilde{c}^I \; \sum_{i \in \parallel_I} u^i \; \; .
\label{ciui}
\end{eqnarray}

We now write down the leading order asymptotic solutions to the
equations of motion and perturbations around them. They will
describe the analogs of the singular solutions and the
asymptotic perturbations around them. We have $\tilde{s} = 0$
and $q^I = - 2 \;$. Equation (\ref{q2i}) can then be satisfied
for all $I$ by choosing $w^I_\pi = w_\pi$ and $w^I = w$ for all
$I \;$, see footnote {\bf \ref{redN}}. Then $\eta^I = \eta$, $\;
\tilde {c}^I = \tilde {c}$, and $\tilde {c}^{0 I} = \tilde
{c}^0$ for all $I$, but $\tilde{c}^{i I}$ do depend on $I$ since
$\tilde{c}^{i I} = \tilde {c}^\parallel \;$ if $\; i \in \;
\parallel_I $ and $\tilde {c}^{i I} = \tilde {c}^\perp \;$
otherwise. Now, using equation (\ref{ciui}) for the $2 2' 5 5' :
(12, 34, 13567, 24567)$ configuration, note that
\begin{equation}\label{sumIiu}
\sum_I \left( \sum_i c^{i I} \; \tilde{u}^i \right) \; = \;
\tilde{c} \; U^c \; ({\cal N} \; - \; 2 \; m ) \; = \; 0 
\end{equation}
since ${\cal N} = 4$ and $m = 2 \;$. \footnote{ \label{spl}
Cancellations of this type do not happen for all intersecting
brane configurations. It happens in the present case, and for
the case where three stacks of two branes intersect (for which
${\cal N} = 3$, $\; n_c = 6$, and $m = 3$), and for the
equivalent U dual versions of these two configurations. Similar
cancellations happen for these two independent configurations in
the cosmological context also. There, the cancellations may be
understood as arising due to the balancing of contraction or
expansion forces applied by the branes on the compcat directions
parallel or transverse to the worlvolume directions \cite
{bdr, bdr2, bdr3}. These cancellations are responsible for the
stabilisation of the compact toroidal directions.} Consider sums
of the type $\sum_I \tilde{c}^{i I} \; X_I \;$. Upon using
$\tilde {c}^\parallel = - \; \tilde{c}^\perp$ and denoting the
$X_I$'s as $(X_2, \; X_{2'}, \; X_5, \; X_{5'}) \;$, such sums
become
\begin{eqnarray} 
\sum_I \tilde{c}^{1 I} \; X_I & = & \tilde{c}^\parallel \; 
\left( X_2 - X_{2'} + X_5 - X_{5'} \right) \nonumber \\
\sum_I \tilde{c}^{2 I} \; X_I & = & \tilde{c}^\parallel \; 
\left( X_2 - X_{2'} - X_5 + X_{5'} \right) \nonumber \\
\sum_I \tilde{c}^{3 I} \; X_I & = & \tilde{c}^\parallel \; 
\left( - X_2 + X_{2'} + X_5 - X_{5'} \right) \nonumber \\
\sum_I \tilde{c}^{4 I} \; X_I & = & \tilde{c}^\parallel \; 
\left( - X_2 + X_{2'} - X_5 + X_{5'} \right) \nonumber \\
\sum_I \tilde{c}^{5, 6, 7 \; \; I} \; X_I & = &
\tilde{c}^\parallel \;
\left( - X_2 - X_{2'} + X_5 + X_{5'} \right) \; \; .
\label{ciXi}
\end{eqnarray}

Consider equation (\ref{q2i}). Upon using equation (\ref{cisi}),
it gives
\[
\sum_{i \in \parallel_I} s^i \; = \; \frac { (1 + w_\pi) \;
\tilde{s}^0 - 2 w_\pi \; (1 - \eta)} {m \; \tilde{c}} \; 
+ \; \frac{S^c}{m} \; \; , 
\]
which implies that the sum $\sum_{i \in \parallel_I} s^i$ must
be same for all $I \;$. Thus, for the $2 2' 5 5' : (12, 34,
13567, 24567)$ configuration, it follows that
\[
s^1 + s^2 = s^3 + s^4 = s^1 + s^3 + s^5 + s^6 + s^7 
= s^2 + s^4 + s^5 + s^6 + s^7
\]
\[
\Longrightarrow \; \; \; 
s^3 = s^2 \;\; , \; \; \;  
s^4 = s^1 \;\; , \; \; \;  
s^5 + s^6 + s^7 = 0 \;\; , \; \; \;  
S^c = 2 \; (s^1 + s^2) \; \; .
\]
It can then be shown that $\tilde{s}^i = s^i + \frac{S^c}{m} \;$
satisfy the relations
\begin{equation}\label{3stilde}
\tilde{s}^3 = \tilde{s}^2 \;\; , \; \; \;  
\tilde{s}^4 = \tilde{s}^1 \;\; , \; \; \;  
\tilde{s}^5 + \tilde{s}^6 + \tilde{s}^7 = \tilde{s}^1 +
\tilde{s}^2 \; \; .
\end{equation}

Consider equation (\ref{00i}) for $\tilde{s}^i \;$: $\alpha \;
\tilde{s}^i = \sum_I \tilde{c}^{i I} R_I \;$. Applying equations
(\ref{ciXi}) now gives
\begin{eqnarray*}
\alpha \; \tilde{s}^1 \; = \; 
\sum_I \tilde{c}^{1 I} R_I & = & \tilde{c}^\parallel \; 
\left( R_2 - R_{2'} + R_5 - R_{5'} \right) \\
\alpha \; \tilde{s}^2 \; = \; 
\sum_I \tilde{c}^{2 I} R_I & = & \tilde{c}^\parallel \; 
\left( R_2 - R_{2'} - R_5 + R_{5'} \right) \\
\alpha \; \tilde{s}^3 \; = \; 
\sum_I \tilde{c}^{3 I} R_I & = & \tilde{c}^\parallel \; 
\left( - R_2 + R_{2'} + R_5 - R_{5'} \right) \\
\alpha \; \tilde{s}^4 \; = \; 
\sum_I \tilde{c}^{4 I} R_I & = & \tilde{c}^\parallel \; 
\left( - R_2 + R_{2'} - R_5 + R_{5'} \right) \\
\alpha \; \tilde{s}^{5,6,7} \; = \; 
\sum_I \tilde{c}^{5,6,7 \; \; I} \; R_I & = &
\tilde{c}^\parallel \;
\left( - R_2 - R_{2'} + R_5 + R_{5'} \right) \; \; . 
\end{eqnarray*}
The three relations (\ref{3stilde}) on the $\tilde{s}^i$ then
imply that (see footnote {\bf \ref{spl}})
\[
R_2 = R_{2'} = R_5 = R_{5'} = \frac{R}{4} 
\; \; , \; \; \; R = \sum_I R_I \; \; , 
\]
\[
\Longrightarrow \; \; \; 
\tilde{s}^i = 0 \; \; \; \Longrightarrow \; \; \; 
s^i = S^c = {\cal B}_0 = {\cal B}_1 = 0 \; \; . 
\] 
Using equations (\ref{00i}) -- (\ref{0sigma}), we then get the
same zeroth order results as for the $(m + 2)$ dimensional stars
where $m = 2$ now. Namely, we get
\begin{eqnarray*}
\tilde{s}^0 & = & \frac {2 \; w_\pi \; (1 - \eta) } {1 + w_\pi}
\; \; \Longrightarrow \; \; \; 
\alpha \; = \; m - 1 + \tilde{s}^0 \; = \; 
\frac {m \; \tilde{c}^0} {1 + w_\pi} \\
& & \\
R & = & \frac {\alpha \; \tilde{s}^0} {\tilde{c}^0} \; = \;
\frac {2 \; m \; w_\pi \; (1 - \eta) } {(1 + w_\pi)^2} \\
& & \\
(m - 1) \; e^{2 \tilde{\lambda}_0} & = & \alpha - \tilde{c} \; R
\; = \; \frac {{\cal D}} {(1 + w_\pi)^2}
\end{eqnarray*}
where ${\cal D} = (m - 1) (1 + w_\pi)^2 + 4 w_\pi (1 - \eta) \;$
and $m = 2 \;$.

\vspace{2ex}

Consider the first order equations of motion (\ref{yi}) and
(\ref{2quad}) -- (\ref{2i}). Their solutions will give the
asymptotic perturbations around the singular solutions.
Equations (\ref{yi}) and (\ref{ciui}) give
\[
w_\pi \; y^I \; = \; - \; (1 + w_\pi) \; \tilde{u}^0 \; 
- \; \tilde{c} \; U^c \; + \; m \; \tilde{c} \; \sum_{i \in
\parallel_I} u^i \; \; . 
\]
Using $\tilde {c}^I = \tilde {c} \;$, $\; \tilde {c}^{0 I} =
\tilde {c}^0 \;$, and ${\cal B}_0 = {\cal B}_1 = 0 \;$, the
equations for $\tilde{u}^0$ and $\tilde{u}$ can be written as
\begin{eqnarray*}
\tilde{u}^0_{\tilde{\sigma}} & = &
\frac{w_\pi}{m} \; \sum_I R_I \; y^I 
+ \left( 2 \tilde{s}^0 + m - 1 \right) \tilde{u} \\ 
& & \nonumber \\
\tilde{u}^0_{\tilde{\sigma}} - \tilde{u}_{\tilde{\sigma}} 
& = & \tilde{c} \; \sum_I R_I \; y^I 
+ 2 \alpha \tilde{u} \\ 
\tilde{u}_{\tilde{\sigma}} & = &
\frac{1}{m} \; \sum_I R_I \; y^I 
- (m - 1) \tilde{u} \\
& & \\
\tilde{u}^0_{\tilde{\sigma} \tilde{\sigma}} + \alpha
\tilde{u}^0_{\tilde{\sigma}} & = & (\tilde{c}^0 - \tilde{s}^0
\tilde{c}) \; \sum_I R_I \; y^I \\
& & \\
\sum_I R_I \; y^I & = & - \; \frac 
{(1 + w_\pi) \; R} {w_\pi} \; \tilde{u}^0 
\; = \; - \; 2 \; (1 - \eta) \; \alpha \; \frac {u^0}
{c^0} \; \; . 
\end{eqnarray*}
The above equation for $\sum_I R_I \; y^I$ follows from $R_I =
\frac {R} {4}$, the expression for $y^I$, equation
(\ref{sumIiu}), and from the results obtained for $R$ at the
zeroth order.

It is easy to see now that these equations for $\tilde{u}^0$ and
$\tilde{u}$ are same as those for $u^0$ and $u$ in the case of
the $(m + 2)$ dimensional stars with $m = 2 \;$. Hence, further
analysis of these equations and the consequent results are also
the same. In particular, $\tilde{u}^0$ and $\tilde{u}$ obey
equation (\ref{*1}). And, their solutions are non oscillatory if
the values of the anisotropy parameter $\eta$ and, equivalently,
of $\frac {w} {w_\pi}$ lie in the ranges given in equation
(\ref{etam+2}). Thus, since $m = 2 \;$, certain amount of
anisotropy, $\eta \ge \frac {7} {8} \;$, is needed to obtain the
non oscillatory behaviour of the perturbations around the
singular solutions in the asymptotic region.

Consider now the equations for $\tilde {u}^i \;$. They are not
needed for present purposes but we analyse them for the sake of
completeness. Since $\tilde {s}^i = 0$ and $R_I = \frac{R}{4}$,
they are given by
\begin{equation}\label{ui2255}
\tilde{u}^i_{\tilde{\sigma} \tilde{\sigma}} + \alpha
\tilde{u}^i_{\tilde{\sigma}} \; = \; \frac {R} {4} \; 
\sum_I \tilde{c}^{i I} \; y^I \; \; . 
\end{equation}
Evaluating the sum $\sum_I \tilde{c}^{i I} \; y^I \;$ using
equations (\ref{ciXi}) gives \footnote{In \cite{k13}, the factor
$\frac {m \; \tilde{c}} {w_\pi}$ appearing below was omitted
inadvertently. Consequently, the equation for $(*_1)$ given
there is incorrect upto this factor in the last term. Equation
(\ref{*12255}) for $(*_1)$ given below is the correct one.}
\[
\sum_I \tilde{c}^{i I} \; = \; 0 
\; \; \; \Longrightarrow \; \; \;
\sum_I \tilde{c}^{i I} \; y^I \; = \; 
\left( \frac {m \; \tilde{c}} {w_\pi} \right)\;
\sum_I \tilde{c}^{i I} \;
\left( \sum_{j \in \parallel_I} u^j \right) \; \; .
\]
For $I : (2, 2', 5, 5')$, the sums $\sum_{j \in \parallel_I}
u^j$ are given by $(u^1 + u^2)$, $\; (u^3 + u^4)$, $\; (u^1 +
u^3 + u^5 + u^6 + u^7)$, and $(u^2 + u^4 + u^5 + u^6 + u^7) \;$.
Using equations (\ref{ciXi}) again gives
\begin{eqnarray} 
\sum_I \tilde{c}^{1 I} \; y^I & = & 2 \; 
\tilde{c}^\parallel \; (u^1 - u^4) \nonumber \\
\sum_I \tilde{c}^{2 I} \; y^I & = & 2 \; 
\tilde{c}^\parallel \; (u^2 - u^3) \nonumber \\
\sum_I \tilde{c}^{3 I} \; y^I & = & 2 \; 
\tilde{c}^\parallel \; (u^3 - u^2) \nonumber \\
\sum_I \tilde{c}^{4 I} \; y^I & = & 2 \; 
\tilde{c}^\parallel \; (u^4 - u^1) \nonumber \\
\sum_I \tilde{c}^{5, 6, 7 \;  \; I} \; y^I & = & 2 \; 
\tilde{c}^\parallel \; (u^5 + u^6 + u^7) \; \; . \label{rhsui}
\end{eqnarray}
It then follows, after some manipulations involving $u^i$s,
$\tilde{u}^i$s, equations (\ref{ui2255}) and (\ref{rhsui}), that
\begin{equation}\label{*12255} 
(*_1)_{\tilde{\sigma} \tilde{\sigma}} + \alpha
(*_1)_{\tilde{\sigma}} + \frac {2 \; R \; \tilde{c}^2} {w_\pi}
\; (*_1) \; = \; 0 
\end{equation}
where $(*_1) = (\tilde{u}^1 - \tilde{u}^4) \;$, $\; (\tilde{u}^2
- \tilde{u}^3) \;$, and $(u^5 + u^6 + u^7) \;$; and 
\begin{equation}\label{*2} 
(*_2)_{\tilde{\sigma} \tilde{\sigma}} + \alpha
(*_2)_{\tilde{\sigma}} \; = \; 0 
\end{equation}
where $(*_2) = (\tilde{u}^1 + \tilde{u}^4) \;$, $\; (\tilde{u}^2
+ \tilde{u}^3) \;$, $\; (\tilde{u}^5 + \tilde{u}^6 - 2
\tilde{u}^7) \;$, and $ (\tilde{u}^5 - 2 \tilde{u}^6 +
\tilde{u}^7) \;$.


\vspace{4ex}

\centerline{\bf 7. Conclusion}

\vspace{2ex}

We now conclude with a summary and a discussion of some of the
issues that require further study.

A brief summary of the present paper is as follows. Unitarity of
evolution in gravitational collapses implies that horizonless
objects must exist which can be macroscopic and must be stable.
In this paper, with such objects in mind, we studied the effects
of anisotropy of pressures on the stability of stars. The stars
are assumed to be static and spherically symmetric in the non
compact space, to have suitable isometries along the compact
directions, and to be made up of constituents with linear
equations of state. Studying the singular solutions and
asymptotic perturbations around them, we obtained the criteria
for the perturbations to be non oscillatory.

We studied stars in four or higher dimensional spacetime with no
compact directions, and also two examples of stars in M theory
made up of stacks of (intersecting) two branes and five branes.
A variety of other examples may also be studied using the
present formulation. We find that non oscillatory perturbations
around the singular solutions are possible if an appropriate
amount of anisotropy is present. The details are given in the
paper.

The behaviour of these perturbations lead to corresponding
asymptotic behaviour in the mass -- central density profile of a
star enclosed in a box of radius $r_* \;$. The non oscillatory
behaviour of the perturbations are likely to indicate stability.
In that case, singular solutions will correspond to stable
configurations, and give the mass -- radius relation $M(r_*)
\sim r_*^{m - 1} \;$, in the limit of large central density or
large radius.

Our results suggest that it may be possible to have stable
horizonless objects in four or any higher dimensions, and that
anisotropic pressures may play a crucial role in ensuring their
stability. Although much remains to be done, it is worth
emphasising that these are important results because they bear
on the horizonless objects which are implied by unitarity in
lieu of black holes, and they point out a necessary ingredient
for their stability. To actually construct such objects,
however, requires detailed understanding of many issues such as
the nature of the constituents and the physical mechanisms that
may provide the required amount of anisotropy. We now discuss
some of these issues which may be studied further.

To show the stability of the equilibrium configurations given by
the mass -- central density profile whose asymptotic behaviour
is monotonic non oscillatory, one may follow Chavanis and show
that the equilibrium configurations in the mass -- central
density profile become unstable beyond the first maximum; then
construct the entire profile and show that it remains monotonic
and increasing in both the non asymptotic and asymptotic
regions. Although desireable, we are unable do any of this since
the detailed properties of the constituents are not known which
cause the required amount of anisotropy. Proving these things
may give valueable insights into the horizonless objects.

One may try to use scalar fields as in \cite {mg} to produce
anisotropy and thereby to construct a stable horizonless
object. The stars constructed in these works are unstable for a
sufficiently high mass. This may be because the scalar field
potentials are not tailored to generate anisotropic linear
equations of state. In the cosmological context, a linear
equation of state can be mimicked using a scalar field with an
exponential potential. One may similarly try to mimic
anisotropic linear equations of state with scalar fields with
appropriate potentials and then study the resulting stars.

There is a vast body of works devoted to construction of
anisotropic stars. A small sample of them is given in \cite
{dg123} -- \cite{hb}. Typically, in these works, it is found
that anisotropy affects stability properties, and that
instability sets in for a sufficiently high mass. The nature of
their ansatzes for anisotropy is very different from ours and,
hence, there seems to be no discernible contradiction between
their results and ours.

There is, however, a distinct possibility that anisotropy of
pressures as found here is a necessary condition for stability
but it may not be sufficient; other ingredient(s) may also be
needed. There are two reasons for entertaining this possibility.
First, in the works \cite{cl} -- \cite{cfv} on horizonless
objects, anisotropy of the pressures was found to be an
important ingredient; but a positive cosmological constant, more
generally matter with negative pressures, residing in the inner
region was also required in an essential way. Second, if there
is a similarity between the singularities in gravitational
collapses and in cosmological big bang/crunch evolutions then
the mechanisms resolving these singularities may also be
expected to be similar. Usually, such mechanisms involve new
types of matter and/or interactions. For example, matter
violating null energy conditions can cause a bounce and resolve
big bang/crunch singularity. Then, going by the similarities,
one may also expect similar ingredients to play a role in
stabilising gravitational collapses. Thus it seems possible that
other ingredient(s), besides anisotropy, may still be needed for
constructing stable horizonless objects. Nevertheless, using the
anisotropy criteria given here, one may try to construct such
objects and see if, and which, further ingredients are needed.

At a technical level, although our formulations included multi
component fluids, we only considered cases where ${\cal N} = 1
\;$, or chose $w^I_\pi = w_\pi$ and $w^I = w$ for all $I$ when
${\cal N} = 4 \;$. It may be worthwhile to investigate
situations where more than one component become crucial and play
a significant role. For example, is it possible that one
component dominates the inner regions and another the outer
regions but such that, together, they lead to stability against
collapse?




\vspace{6ex}


\end{document}